\documentclass{article}
\usepackage{setspace}

\usepackage{PRIMEarxiv}

\usepackage[utf8]{inputenc} 
\usepackage[T1]{fontenc}    
\usepackage{url}            
\usepackage{booktabs}       
\usepackage{amsfonts}       
\usepackage{nicefrac}       
\usepackage{microtype}      
\usepackage{lipsum}
\usepackage{fancyhdr}       
\usepackage{graphicx}       
\usepackage{bm}
\usepackage{multirow}
\usepackage{adjustbox}
\usepackage{float}
\usepackage{parskip}
\usepackage{algorithm,algpseudocode}

\graphicspath{{media/}}     
\RequirePackage{amsthm,amsmath,amsfonts,amssymb}
\RequirePackage[round,authoryear]{natbib}
\RequirePackage[colorlinks,citecolor=blue,urlcolor=blue,linkcolor=blue]{hyperref}

\title{Emulation with uncertainty quantification of regional sea-level change caused by the Antarctic Ice Sheet}

\author{
  Myungsoo Yoo\thanks{Corresponding author} \\
  University of Missouri \\
  \texttt{mym4v@mail.missouri.edu} \\
   \And
  Giri Gopalan\\
  Los Alamos National Laboratory\\
  \texttt{ggopalan@lanl.gov} \\
   \And
  Matthew J. Hoffman\\
  Los Alamos National Laboratory\\
  \texttt{mhoffman@lanl.gov} \\
   \And
  Sophie Coulson\\
  University of New Hampshire\\
  \texttt{sophie.coulson@unh.edu} \\
   \And
  Holly Kyeore Han\\
  Los Alamos National Laboratory \\(now at Jet Propulsion Laboratory)\\
  \texttt{kyeore.han@jpl.nasa.gov} \\
   \And
  Christopher K. Wikle\\
  University of Missouri \\
  \texttt{wiklec@missouri.edu} \\
   \And
  Trevor Hillebrand\\
  Los Alamos National Laboratory \\
  \texttt{trhille@lanl.gov} \\
}

\begin{document}
\maketitle

\begin{abstract}
Projecting regional sea-level change under various climate-change scenarios typically involves running forward simulations of the Earth's gravitational, rotational and deformational (GRD) response to ice-mass change, which requires substantial computational cost if applied to probabilistic frameworks requiring thousands to millions of samples. Here we build emulators of regional sea-level change at 27 coastal locations, due to the GRD effects associated with future Antarctic Ice Sheet mass change over the 21st century. The emulators are evaluated against a numerical sea-level model applied to an ensemble of ice-sheet model simulations of the Antarctic Ice Sheet through 2100.  We build a physics-based emulator using a recent sensitivity kernel approach and compare it to machine learning based emulators (neural network and conditional variational autoencoder methods). In order to quantify uncertainty, we derive well-calibrated prediction intervals for regional sea-level change via split-conformal inference and linear regression, and show that Monte Carlo dropout does not yield well-calibrated uncertainties in this instance. We also demonstrate substantial gains in computational efficiency using both the physics-based emulator and neural networks in comparison to the numerical model for the complete regional sea-level solution.  Overall, we find the physics-based emulator modestly outperforms the machine learning emulators for this problem.
\end{abstract}

\section*{Plain Language Summary}
Sea-level change varies regionally because changes in the weight of ice sheets and glaciers deform the Earth’s crust and modify the Earth’s gravitational field and axis of rotation. While numerical sea-level models are widely used to project sea-level change from these effects, they require thousands of evaluations to quantify uncertainty in the projections. To accelerate this process, we evaluate the feasibility of different simplified models that trade some accuracy for substantial computational speedup.  We consider both machine learning methods and a reduced-physics approach and apply these ``emulators'' to 27 major coastal locations. The emulators demonstrate high prediction accuracy and are between 160 - 240 times faster to run than the numerical sea-level model. We then quantify the uncertainty in regional sea-level change introduced by the emulators themselves using two techniques. Based on this, we find that the reduced-physics approach is most accurate, fast to evaluate, is easy to generate, and produces accurate uncertainties when coupled with the uncertainty quantification techniques aforementioned.

\section{Introduction}
\label{sec:intro}
Coastal communities and infrastructure are under severe threat from sea-level rise caused by climate change \citep{foxkemper2021ocean, Hauer2020a}. Global mean sea level (GMSL) rose by 0.20 m over the last century, and the most recent Intergovernmental Panel on Climate Change assessment report (\cite{foxkemper2021ocean}) indicates that it is "virtually certain" that GMSL will continue to rise through 2100. Continued sea-level rise is expected regardless of anthropogenic greenhouse gas emissions scenario due to the likelihood of all major processes contributing to GMSL continuing to operate.  
Sea-level change varies spatially, and relative sea level (RSL, i.e., thickness of the ocean) can differ by up to a factor of two relative to GMSL across regions of the globe \citep{foxkemper2021ocean, Kopp2015}.  Variations in RSL are caused by steric effects (changes in ocean density and salinity) and dynamic sea level changes, as well as  gravitational, rotational, and deformational (GRD) effects \citep{Slangen2014, Kopp2015, Gregory2019, Hamlington2020, foxkemper2021ocean}. Unlike the other two factors, GRD effects are induced by changes in surface mass loading such as ice sheets and ocean, glaciers, and terrestrial water storage (reservoirs, lakes, groundwater). That is, mass redistribution on the Earth's surface deforms the solid Earth, perturbs the Earth's gravitational field, and shifts the rotation axis of the Earth. In addition, the solid Earth deforms over time in response to the changing surface loads. These complex processes yield spatially and temporally varying sea-level changes referred to as "sea-level fingerprints" unique to the geometry of ice mass change \citep{Farrell1976, Mitrovica2011}. 

Uncertainty in future sea-level change is dominated by changes in the Antarctic Ice Sheet (AIS) \citep{Kopp2017, Bakker2017, Edwards2021, foxkemper2021ocean}, because it contains large sectors where ice is grounded below sea level and potentially susceptible to rapid retreat through marine ice-sheet instability \citep{Schoof2007}.
Projections of future AIS contribution to sea level are improving through coordinated community model intercomparison efforts \citep{Nowicki2020, Seroussi2020, Edwards2021}, as well as large ensembles of AIS simulations from single models that quantify uncertainties \citep{Bulthuis2019, Berdahl2023, Coulon2024}.
These studies indicate a wide range of potential future outcomes for AIS through 2100 and beyond.  Further, recent studies have shown that the variable geometry of projected AIS mass changes has a significant impact on future RSL due to GRD effects \citep{Roffman2023, Cederberg2023}.

Because sea-level change has large uncertainty and also varies regionally, probabilistic projections of RSL at cities of interest that incorporate the full range of AIS projections are critical for assessing future climate impacts and risk.
These ice-sheet projection ensembles have been used in probabilistic frameworks for analyzing RSL that use Monte Carlo sampling with large numbers of evaluations (order $10^5-10^6$) to produce probability distribution functions of regional sea level at individual cities of interest \citep{Jevrejeva2018, Kopp2023}.  However, these frameworks have so far assumed static sea-level fingerprints calculated by only accounting for the total ice thickness changes for each ice sheet and glacier region \citep{Kopp2023}.   These projections would be improved by considering temporally evolving sea-level fingerprints. Because probabilistic sea-level change frameworks require evaluation of tens of thousands of samples \citep{Wong2017, Kopp2023}, they typically rely on fast-to-evaluate emulators or parameterizations of complex models and processes that include an acceptable trade-off between accuracy and computational cost.  Our goal in this work is to evaluate two methods for introducing sample-specific sea-level fingerprints into the probabilistic frameworks that produce RSL distributions for individual cities: emulation of a sea-level model using 1. machine learning, and 2. sensitivity kernels \citep{Al-Attar2024} derived from an adjoint method for a simplified version of the sea-level model.

An emulator is a surrogate model that mimics the underlying simulator (such as the sea-level model in this study) and is computationally efficient, serving as an alternative to the model it replicates. The significance of an emulator becomes particularly evident when the simulator is computationally too expensive and when the emulator can effectively quantify uncertainties. Because of the complex relationships between mass loading changes and sea-level fingerprints \citep{Mitrovica2011, Roffman2023, Cederberg2023} and costs of calculating time-evolving sea-level fingerprints using sea-level models, machine learning (ML) provides one promising avenue for creating computationally inexpensive emulators of sea-level fingerprints evaluated at locations of interest. 

In essence, ML refers to various computational approaches for approximating the underlying function that encapsulates an input-output relationship. Recent work has demonstrated the potential of a graph-based spherical convolutional neural network to emulate a sea-level model applied to ice-age glacial isostatic adjustment over the last deglaciation and achieved a 100--1000$\times$ speedup \citep{lin2023georgia}.  Another study used an artificial neural network to develop an emulator for the effect of lateral variations in Earth rheological structure that have often been left out of sea-level models \citep{Love2023}.  ML has also been applied to the ice-sheet model projections that form the inputs to sea-level models.  \citet{VanKatwyk2023} created an emulator for the contribution from the AIS to GMSL by training a neural network on an ensemble of ice-sheet model experiments.  \citet{Rohmer2022} used a similar ensemble, in this case for the Greenland Ice Sheet, in conjunction with an ML-based attribution approach to quantify the influence of different ice-sheet modeling decisions to projections of future ice-sheet contribution to GMSL.  ML methods have also been utilized to predict complex regional sea level signals from observed historical sea-level change datasets at month to year timescales \citep[e.g.][]{Tur2021, Ayinde2023, Nieves2021, Guillou2021, Altunkaynak2021, Song2021} and applied to various individual components of the ice-sheet models \citep[e.g.][]{Brinkerhoff2021, Jouvet2022, Jouvet2023, He2023, Riel2021, Riel2023, Rosier2023, Verjans2024}.  However, the use of ML to accelerate probabilistic projections of regional sea-level calculations for individual cities has yet to be attempted.  Of existing work, the sea-level model emulation of \citet{lin2023georgia} is the closest analog, but here we choose to focus on a simpler time-independent problem for near-term projections with a small number of target locations, as opposed to the more complex time-dependent and spatial problem addressed by \citet{lin2023georgia}.

The second approach we evaluate employs a sensitivity kernel for each location of interest (city) against which fields of ice mass change can be multiplied to derive RSL for that location without needing to re-solve the sea-level equation \citep{Al-Attar2024}.  The sensitivity kernel approach exploits recently established reciprocity theorems for the elastic sea-level equation that show that the RSL at one location due to the change in mass load (e.g., from an ice sheet) at another location is equal if the loading and RSL evaluation points are reversed \citep{Al-Attar2024}. The kernel approach requires some minor limiting assumptions (elastic rheology; fixed shoreline, including ice-sheet grounding lines; inability to self-consistently evaluate the ice-sheet flotation condition) be applied to the sea-level equation and must be calculated separately for each location of interest.  These restrictions motivate the comparison with ML emulation as a potential alternative.

In this study, we compare neural-network ML emulation approaches against the kernel sensitivity approach for the purpose of rapid evaluation of many samples at a modest number of geographic locations. We develop multiple neural-network ML emulators for the established sea-level equation solver of \citet{Han2022} for predicting RSL changes caused by changes in the mass of AIS. The emulators are trained on a set of sea-level model simulations forced with a recent ensemble of ice-sheet model projections of AIS through 2100 (ISMIP6, Serrousi et al., 2020).  Because we are focused on methods for producing RSL distributions for individual cities, we train the emulators to consider the locations of 27 major coastal cities; note that the neural-network emulators jointly predict RSL changes for all 27 cities at once, whereas the kernel sensitivity approach has a distinct kernel for each of the 27 locations. We begin by comparing different neural-network based emulators, as well as the kernel sensitivity method, and evaluate their strengths and weaknesses for this application. We then quantify uncertainty in emulator predictions by generating prediction intervals using both split-conformal inference \citep{lei2018distribution} and simple linear regression, a method that has been used for calibrating numerical forecasting models \citep{glahn1972use, glahn2009mos,lovegrove2023improving}. We also compare Monte Carlo dropout \citep{dropout}, a commonly-used method for uncertainty quantification in neural networks. Next, we apply the emulators to generate probabilistic projections of RSL change caused by AIS mass changes by generating an ensemble of artificial ice-sheet projections.  Finally, we assess the relative cost and accuracy of the ML emulation and kernel sensitivity approaches for this application.

\section{Data description}
Training the ML emulators required datasets of simulated ice-sheet mass change projections and the corresponding regional sea-level fingerprints, which were divided into training, validation and test groups. The following sections describe the generation of these datasets in detail.

\subsection{Ice-sheet mass change dataset}
\label{S:ISMIP6}
The input dataset for our workflow was an ensemble of future AIS mass change calculated (i.e., simulated) by ice-sheet models.  For this, we adopted the results of the AIS experiments from the Ice Sheet Model Intercomparison Project for Coupled Model Intercomparison Project Phase 6 (ISMIP6) \citep{Seroussi2020}. The ISMIP6-Antarctica protocol specified standard experiment configurations for ice-sheet models to generate projections of AIS from 2015 through 2100 \citep{Nowicki2020}. The ice-sheet model simulations were forced by atmosphere and ocean output from six different Coupled Model Intercomparison Project Phase 5 (CMIP5) climate models for the high greenhouse gas emissions RCP8.5 scenario and two climate models for the low emissions RCP2.6 scenario.  In addition to testing different climate forcings, different experiments tested different representations of ice-shelf basal melt and the potential for removal of ice-shelf regions through hydrofracture. In total, there were 21 experiments defined; here we select eight experiments for which standardized protocols were prescribed. The experiments were executed by 13 modeling groups around the world, and the results were analyzed on a common reference grid. Each modeling group also performed a control run through 2100 using steady constant climate forcing to quantify model drift. The ensemble represents a wide variety of ice-sheet model complexity, grid resolution, and model characteristics. Notably, none of the ice-sheet models in the ISMIP6 ensemble included GRD effects.

To prepare these ice-sheet projection data as input to a sea-level model, we performed a number of processing steps. First, we calculated changes in the ISMIP6 spatial field ``land\_ice\_thickness'' from its initial value.   
Because there is a large drift in the control run of many simulations \citep{Aschwanden2021}, we removed the thickness change in the control run for each model from each projection, as was done in the ISMIP6 analysis \citep{Seroussi2020}. This step is particularly important for modeling regional sea-level changes, as some portion of ice thickness change from raw projections will be related to the bias inherent to each ice-sheet model (i.e., model drift), which we do not want to feed into our sea-level fingerprint calculation.
While ISMIP6 output is posted annually, we subsampled it to 5 year intervals because annual changes are relatively small, leading to 17 time levels of ice thickness change.
Finally, we remapped the ice thickness change fields from the 8 km polar stereographic projection to the $512\times1024$ latitude/longitude Gaussian grid used by the sea-level model using a conservative remapping method to ensure total AIS mass change is unaltered by the remapping.  

A total of 194 AIS projections were generated for ISMIP6 across the different experiments and models used, and here we restrict ourselves to 71 projections derived from the eight experiments with standardized protocol (experiments 5-10 and 11-12), eliminating a small number of projections due to processing errors related to issues with data format conventions.  Accounting for the 17 time levels retained for each projection, we had a total of 1207 fields of ice thickness change in our input dataset used for training, validation, and testing of ML emulators.

\subsection{Sea-level dataset: Calculation of sea-level fingerprints}

We generated sea-level fingerprints associated with different patterns of AIS mass loss using the sea-level model described by \citet{Han2022}, an updated version of sea-level models described in \citet{Kendall2005} and \citet{Gomez2010a}. The model computes barystatic sea-level changes and associated GRD effects caused by land ice mass changes by solving the static sea-level equation \citep{Farrell1976,Mitrovica2003} using the pseudo-spectral algorithm described in detail in \citet{Kendall2005}, with truncation up to spherical harmonic degree and order 512. The sea-level framework accounts for shoreline migration \citep{Milne_Mitrovica_1998}, including inundation of presently glaciated marine basins as the extent of the grounded ice sheet retreats, as well as perturbations in the Earth's rotation axis \citep{Mitrovica2003, Mitrovica2005}.  Our simulations include a check for floating ice by evaluating the flotation criterion iteratively as RSL is solved.  The flotation criterion assumes an ice density of 910 kg m$^{-3}$ and a water (ice melt and ocean) density of 1000 kg m$^{-3}$.  Note that because we are interested in projections of future sea-level change, there are not observational data to evaluate our sea-level change calculations against and we assume that the calculations by the \citet{Han2022} model are sufficiently accurate.

In addition to the projected patterns of AIS ice mass loss, the simulation requires a model for the rheological structure of the solid Earth. We adopt a radially symmetric elastic model for the Earth with elastic and density properties derived from the seismic PREM model \citep{Dziewonski1981}. 
We also evaluated simulations with viscoelastic Earth structure using the depth-varying viscosity of the Earth given by the viscosity profile VM7 \citep{Roy2017, Roy2018}. For the 85 year time-frame of the ISMIP6 experiments the contribution from viscous deformation was negligible.
We note that these viscosity values represent a global average; while appropriate for East Antarctica and our 27 far field coastal locations, this Earth structure does not accurately capture the Earth structure in West Antarctica where upper mantle viscosity is thought to be 2-3 orders of magnitude lower than the global average \citep{Barletta2018, Nield2014}. 
We note that while it is well-established that the Earth structure in Antarctica shows strong lateral heterogeneity \citep{Lloyd2020} likely having an effect on projections of local sea-level change, the sensitivity of far-field sea level to the 1D versus 3D Earth structure only becomes large enough to consider in a long-term (multi-centennial or longer) scale projection \citep{Pan2021, Powell2022}, which is beyond the timescale we consider in this work.  We conducted additional sensitivity tests using a lower mantle viscosity appropriate for West Antarctica and confirmed that results differed only by a few percent from the elastic-Earth configuration outside the Antarctic for these 85-year simulations, justifying our use of an elastic-only Earth structure. 

\begin{figure}[t]
\caption{Illustration of RSL fingerprints for given AIS thickness changes for two samples representative of small and large AIS mass change.
a) Modeled AIS grounded-ice thickness change between 2015 and 2100 for sample from training dataset with third smallest total AIS mass change.
b) Calculated RSL change fingerprint for AIS mass loss in a.
c) Modeled AIS grounded-ice thickness change between 2015 and 2100 for sample from training dataset with third largest total AIS mass change.
d) Calculated RSL change fingerprint for AIS mass loss in c.
Note different colorbar ranges for each plot and that values exist beyond the ranges shown.
The 27 long-term tide gauge sites selected by \citet{Meyssignac2017a} that we used as emulation targets are shown with black dots in b and d.
}
\label{fig:sim_example}
\centering
\includegraphics[width=1.0\linewidth]{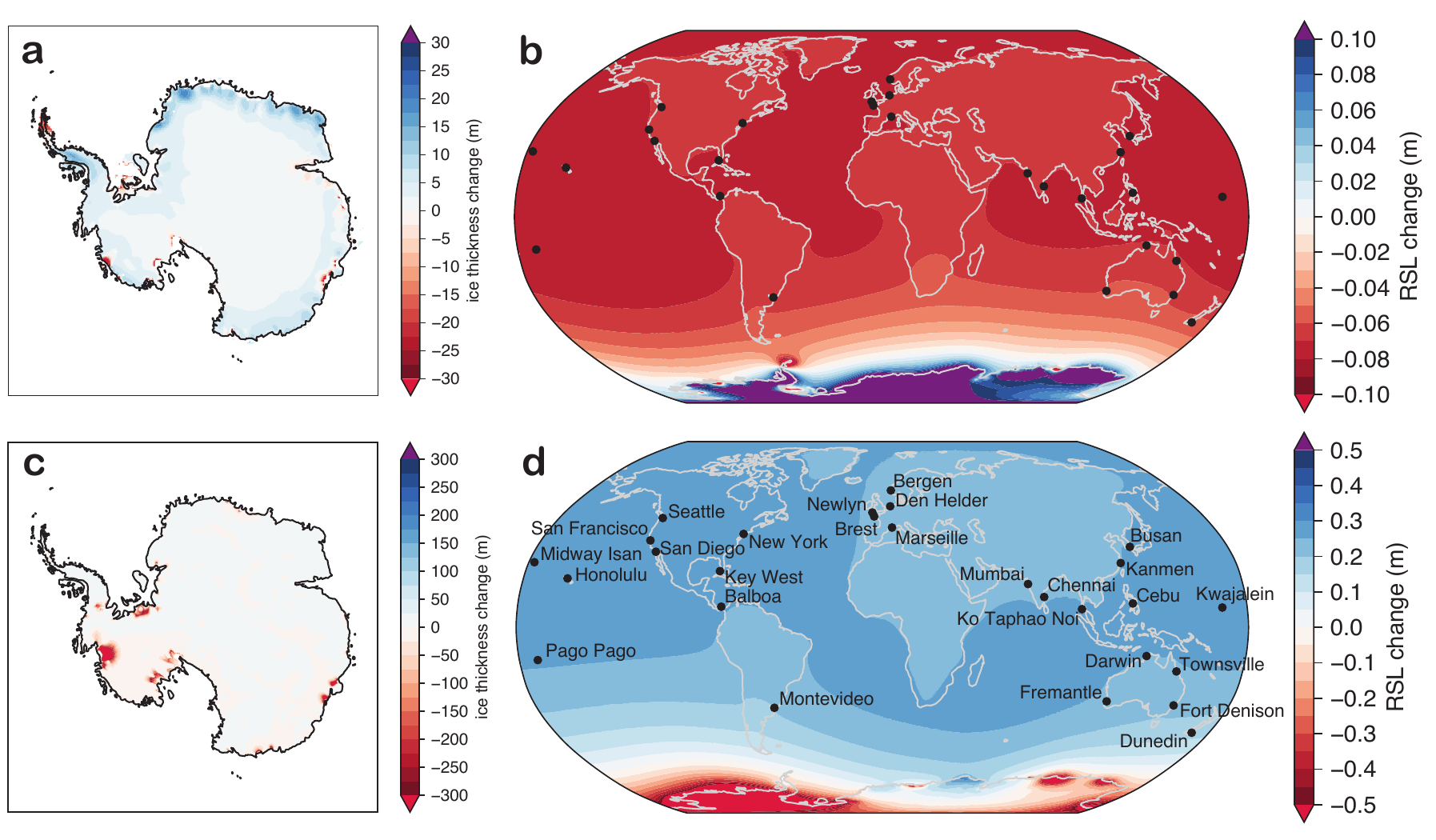}
\end{figure}

Since the sea-level model runs on a global grid but the original ISMIP6-AIS datasets were provided on a local grid, we adopt topography from the ETOPO2
dataset \citep{NOAA2022} for the region outside of the ice-sheet model domain. With the pre-processed ice-sheet thickness change time-series and the Earth structure described above, we perform a suite of sea-level simulations with 5-year time intervals and generate sea-level change output datasets for emulator training, validation, and testing. Examples of the maps of sea-level change associated with AIS thickness changes are illustrated in Figure \ref{fig:sim_example}.  Note that while the 17 time levels evaluated are sequential in time, the elastic GRD effects considered are instantaneous and there is no time-dependence in the model solutions.  That is, the RSL field is calculated from a given ice thickness change field with no dependence on ice thickness change history.

We initially attempted to emulate the entire $512\times1,024=524,288$ dimension output space of the sea-level model (per time step), with the best-performing model using an autoencoder \citep{autoencoder} for dimension reduction on inputs and outputs and a feed-forward neural network linking between reduced-dimension spaces. Our initial results varied considerably in quality -- while some of the entire fields in the test set could be emulated reliably, about 10\% of the global test fields were poorly emulated (normalized root mean squared error exceeding 0.11). Moreover, the latitude-longitude coordinates of the sea-level model spatial output added complexity in accurately representing a spherical surface in two dimensions \citep{lin2023georgia}. Given our primary goal of emulating RSL at a set of selected cities, the global RSL field was not strictly necessary.  Therefore, to improve the accuracy of the ML emulators, we chose to only evaluate RSL change at the 27 long-term tide gauge sites identified by \citet{Meyssignac2017a} (Fig.~\ref{fig:sim_example}c), which come from the Permanent Service for Mean Sea Level \citep{Holgate2013}.

\subsection{Partitioning of training, validation, and test datasets}

The entire dataset of ISMIP6 AIS mass change fields, along with their corresponding numerically solved sea-level change outputs, was split into training, validation, and test datasets, as is commonly done in ML (\cite{hastie2009elements}), with 80\% used for training, 10\% used for validation, and 10\% used for a hold-out test set. As mentioned above, this constitutes 1,207 pairs of AIS mass change and global relative sea-level change fields across the different ice-sheet models, experiments, and time levels.

In order to ensure representation of each year across the three data splits, the input-output data pairs from the 71 ice-sheet simulations per each year were randomly selected to be in the training, validation, and test sets; that is, we stratify by year when constructing the training, validation, and test splits. Of the 71 total ice-sheet simulations for a given year, 57 were randomly selected to be in the training set, 7 were randomly selected to be in the validation set, and 7 were randomly selected to be in the test set, yielding 80\% for training, 10\% for validation, and 10 \% for test. The final split of data consisted of 969 in training, 119 in validation, and 119 in testing, where the inputs of each split were standardized to have 0 mean and unit variance. The length of the flattened input vector for each time step of each simulation was 579,121, and the output vector was the 27 long-term tide gauge locations.

\section{ML models}
\label{sec3}
We constructed machine learning (ML) models that take as input a vector of ice thickness changes across Antarctica derived from ISMIP-6 simulations and outputs a prediction of the simulation of resultant RSL change at 27 long-term tide gauge locations (Fig.~\ref{fig:sim_example}c,d). That is, in our RSL change emulator, the input is the flattened vector of ice-thickness changes, and the output is a prediction of the resultant RSL change for the 27 long-term tide gauge locations. Ideally, the ML model should accurately predict simulated RSL change values, be computationally more affordable than running the full numerical simulation, and also enable statistically valid uncertainty quantification of predictions. In what follows, we describe neural network ML implementations that are constructed and compared: a feedforward neural network (NN) and a conditional variational autoencoder (CVAE). We choose neural networks due to their prevalence in many widely-used modern ML implementations, including transformers in large language and vision models \citep{vaswani2017attention}, for physics-informed ML \citep{raissi2019physics}, and in generative artificial intelligence \citep{achiam2023gpt}. We discuss other possibilities at the end of the section.

\subsection{Feedforward neural network (NN)}
A feedforward neural network (NN) is a fundamental component of deep neural networks. The primary objective of an NN is to learn (or approximate) a mapping $\bm{y}=\bm{f}(\bm{x};\bm{\theta})$ from input $\bm{x}$ to output $\bm{y}$, accomplished through the composition of functions governed by parameters $\bm{\theta}$ that consist of weights and bias terms. The approximation task is equivalent to determining the best estimate $\hat{\bm{\theta}}$ that minimizes a loss function (e.g., mean-squared error) using a training set, which is typically updated iteratively with backpropagation \citep{Rumelhart}. 

It is common to use well-known optimization routines such as stochastic gradient descent (SGD) \citep{sgd} and ADAM \citep{adam} for fitting neural networks. 
A feedforward neural network (NN) consists of an input layer, $L$ hidden layers, and an output layer, where $L$ denotes the number of hidden layers. $\ell^{th}$ hidden layer is equipped with $d_{\ell}$ neurons, which serve as the dimensions of hidden states. 

The layers' units (neurons) are interconnected through a linear transformation and an activation function, denoted as $g(\bm{x}^{\top}\bm{w}+\bm{b}$), where $\bm{w}$, $\bm{b}$, and $g(\cdot)$ correspond to the weight, bias, and an activation function, respectively. A sequence of these operations across each layer can yield a high-quality approximation of complex functions. The potential for such approximation is enhanced by the activation function, such as the rectified linear unit (RELU), which introduces non-linearity. In addition to its rich approximating capabilities, user-friendly software, such as PyTorch by \citet{pytorch}, has contributed to the popularity of NNs. For a more detailed introduction to NNs, refer to \citet{goodfellow}.

In our implementation, we utilize a neural network architecture comprising $L=4$ hidden layers. Due to the multiple layers, we consider this to be a deep neural network \citep{goodfellow}. We use the RELU activation function between layers, while a linear activation function is applied between the last hidden layer and the output layer. The number of neurons within each layer is $(d_{1},d_2,d_3,d_4)=(3200, 1600, 800, 400)$. Using root mean squared error (RMSE) as the loss function, we train the model with the SGD optimizer with a learning rate of 0.001. Throughout the training process, we monitored the convergence of the loss on the validation set. To prevent overfitting, we also introduce dropout \citep{dropout} with a probability 0.3 between layers. We implement the NN using Pytorch version 2.0.1 with Python version 3.9.12. 

\subsection{Conditional variational autoencoder (CVAE)}
\begin{figure}[t]
\caption{Illustration of the CVAE architecture. The encoder, in the left-most box, is a feed-forward neural network that maps to the latent space in the middle of the diagram. This latent space is lower dimensional than the input space. For generating output, a latent multivariate normal vector, depicted with a density plot, is sampled and concatenated with conditional information, and then it is subsequently passed through the decoded network to produce a prediction. Note that this a generative model as multiple outputs can be generated with the same conditional information so long as new latent vectors are sampled.}
\label{fig2_cvae}
\centering
\includegraphics[width=1\linewidth]{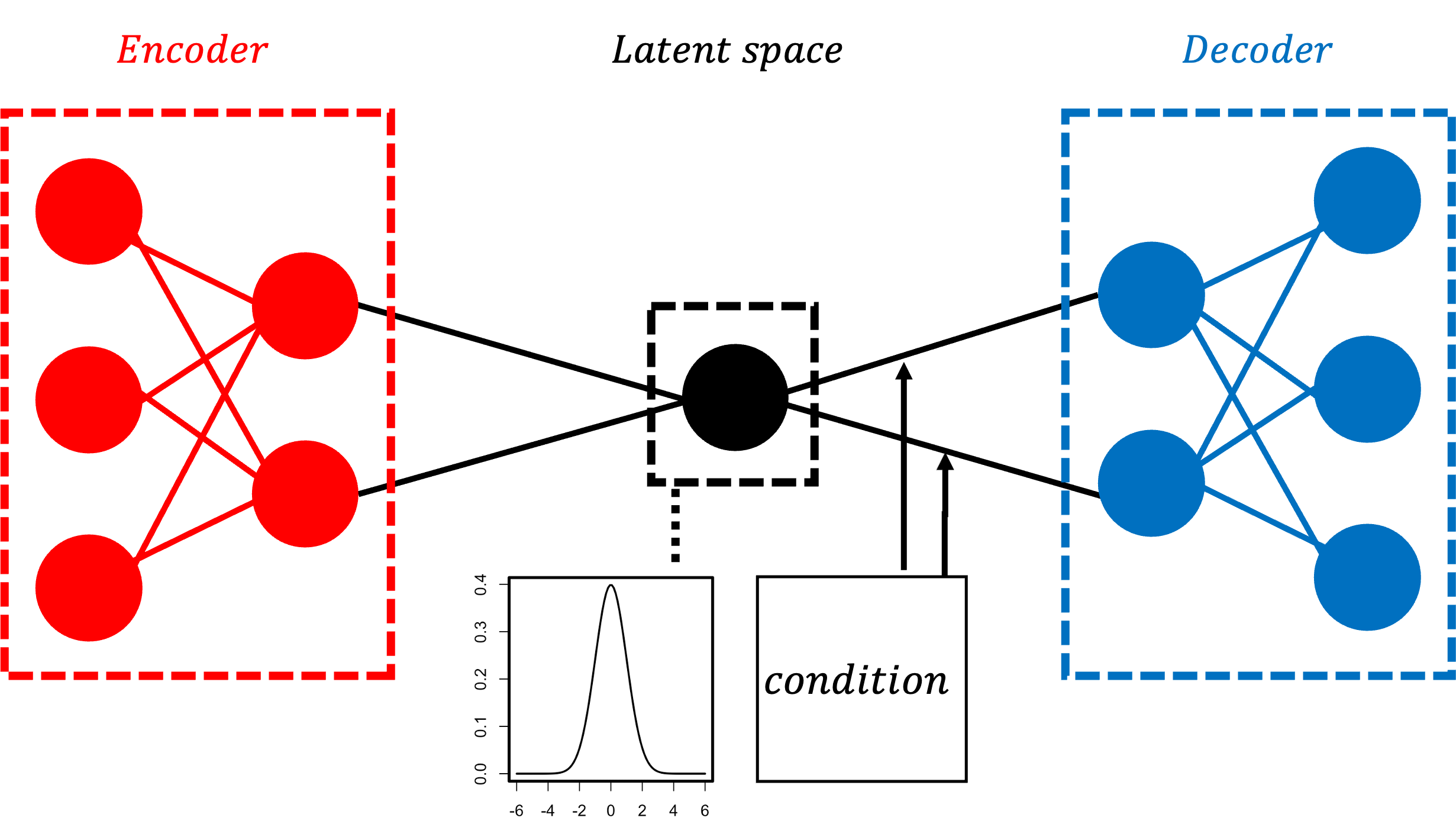}
\end{figure}
The CVAE is a widely-used modern neural network architecture that couples an encoder to a decoder, with the intermediate layer between them representing a latent, lower-dimensional space. The CVAE is appealing and can be effective in constructing generative models (e.g., a molecular generative model \citep{lim} and zero-shot learning generative model \citep{ Mishra_2018_CVPR_Workshops}). We consider a CVAE in addition to the feed-forward NN in order to attempt a generative approach for the sea-level change emulator, since by sampling from the appropriate latent space, one can generate samples of predictions for a given input as opposed to a single prediction; indeed this approach serves as a precursor for generative artificial intelligence (AI) approaches such as diffusion models \citep{ho2020denoising}.

The origin of the CVAE can be traced back to the autoencoder (AE) \citep{autoencoder} and the variational autoencoder (VAE) \citep{vae}. In general, an AE is trained to reproduce the original input, thereby learning a reduced-dimension latent representation that captures much of the information of the higher dimensional input data. Although the architecture of the VAE is similar to the AE, the VAE assumes that the reduced-dimensional latent space follows a certain distribution, often a multivariate normal with 0 mean and identity covariance matrix. The key advantage of the VAE over the AE is that it can serve as a generative model by drawing a sample from the distribution of the latent, intermediate layer.

While both the CVAE and the VAE are similar, the CVAE possesses an additional feature: the ability to instruct the model to predict the output based on conditional information. In essence, a prediction for a new input is determined by concatenating a random draw from the latent space with conditional information, which in our setting is the vector of ice-thickness changes. This concatenated vector, which combines conditional information with latent information, is passed through the decoder to make a prediction. For a visual representation, see Figure \ref{fig2_cvae}; more details are described in the tutorial of \cite{doersch2016tutorial}.

Our implementation of the CVAE for the sea-level change emulator takes the entire vector of ice-thickness changes as conditional information, concatenates it with a latent draw, and passes the concatenated vector through the decoder to make a prediction. Specifically, the encoder comprises $L=4$ hidden layers with neurons ($d_1$, $d_2$, $d_3$, $d_4$) = (3200, 1600, 800, 400), and it outputs 200 pairs of parameters ($\mu_i$, $log \sigma^2_i$), where $i=1,...,200$. These parameters correspond to the parameters of the \textit{posterior} intermediate layer distribution $P(\bm{Z}|\bm{X})$, which is assumed to follow a multivariate normal distribution with a mean vector $\bm{\mu}=(\mu_1,...,\mu_{200})$ and a diagonal covariance matrix $diag(\sigma^2_1,...,\sigma^2_{200})$. Then, the decoder takes as input the concatenated vector of the ice-sheet thickness vector (i.e., the conditional information) and the transformed samples from the \textit{prior} distribution $P(\bm{Z})$ for the intermediate layer. Assuming $P(\bm{Z})$ follows a multivariate normal distribution with 0 mean and identity covariance matrix, the transformation is achieved through re-parametrization $Z_i|\bm{X}=\mu_i+\sigma_i Z_i$ \citep{vae}. Similar to the encoder, the decoder also consists of $L=4$ hidden layers with the same number of neurons as the encoder. It outputs the predicted sea-level change at 27 long-term tide gauge locations. Because the CVAE is predicting sea-level change values as opposed to reconstructing its own input, this is a modification of the usual AE approach.

The loss function used to train is a sum of two terms: the first is a KL-divergence between the variational approximation of \textit{posterior} latent distribution $P(\bm{Z}|\bm{X})$ and \textit{prior} distribution $P(Z)$, which can be thought of as a regularization penalty. The second is a mean squared error term for predictions from the decoder. In our implementation, we draw 500 samples from the latent distribution $P(\bm{Z})$ during the prediction generation process, and the final predicted value is obtained by calculating the mean of these 500 predictions. The CVAE is implemented using PyTorch version 2.0.1 with Python version 3.9.17.

\subsection{Other ML options}
We considered additional ML models besides neural networks as baseline comparisons. These baseline comparisons are a stationary Gaussian process with a squared-exponential covariance function (GP) and a random forest (RF). In our preliminary work, we found that the neural-network emulators performed just as accurately or more accurately than the baselines aforementioned. Additionally, the Gaussian process and random-forest emulators used an additional dimensionality reduction step to allow for scalability, whereas the neural networks are able to handle the order 100,000 dimensional input without a preliminary dimension reduction procedure. The use of dimensionality reduction can be accomplished in a number of ways (e.g., principal components as we implemented or random subspace projections), and so an additional dimension-reduction step introduces a conflating factor in evaluating performance of the GP and RF models. These considerations, coupled with the fact that neural networks are the predominant ML model of choice for modern AI applications, led us to proceed with only the neural-network emulators in comparing to the physics-based kernel sensitivity method, which is next described. Nonetheless, a description and review of the GP and RF approaches is presented in the Appendix. 

\section{Kernel sensitivity method}
\label{kernel}

Although applications of the sea-level equation have been long-established for predicting RSL caused by a specified change in land-ice thickness \citep[e.g.][]{Farrell1976, Kendall2005, Mitrovica2003}, such applications require re-solving the sea-level equation for each unique spatial pattern of land-ice mass change.  In contrast, in recent years, multiple formulations have been made that invert the relationship between land-ice mass change and RSL to produce sensitivity maps of RSL at a given location to the spatial patterns of ice thickness change \citep{Larour2017, Crawford2018, Mitrovica2018, Al-Attar2024}.  These sensitivity kernel approaches are attractive for the problem addressed in this study of evaluating many realizations of the impact of ice-sheet thickness change on RSL at a modest number of coastal locations; by calculating the sensitivity kernels once for each coastal site of interest, the RSL at each site due to any pattern of ice-sheet thickness changes can subsequently be evaluated for the computational cost of a single vector multiplication.

Of the RSL sensitivity-kernel methods that have been described, we selected the recent method of \citet{Al-Attar2024} for this application.  It is restricted to the simpler elastic fingerprint problem we consider here, ignoring the complexities of viscoelastic sensitivity kernels \citep[c.f.][]{Crawford2018}.  It includes the rotational feedbacks that are important for AIS mass changes \citep{Cederberg2023, Roffman2023}.  Lastly, it is easy to implement with any elastic fingerprint model and does not require automatic differentiation or specialized model implementation.  

Although the sensitivity-kernel method is attractive for our application, it includes some simplifications relative to the full sea-level model used. 
In addition to the limitation of elastic rheology, the \citet{Al-Attar2024} method has the restriction of requiring a fixed shoreline position, which renders the elastic fingerprint equation self-adjoint and thus possible to generate sensitivity kernels with a standard forward model.  For the sea-level changes ($\mathcal O (10^{-2})$ m)  from the ISMIP6 AIS ensemble for 2015-2100, shoreline migration from inundation of coastal areas is expected to be negligible.  However, shoreline migration due to the retreat of grounded marine ice sheets will not be, and must be treated approximately in a fixed shoreline model.  A  difference between the \citet{Han2022} and \citet{Al-Attar2024} models is that the \citet{Han2022} model considers the full ice thickness of marine ice sheets and iteratively evaluates the flotation condition as part of the RSL solution.  Additionally, it has an evolving ocean mask that adjusts as grounded ice retreats from or advances into a grid cell, allowing grid cells to dynamically switch between being occupied by ice or ocean (as described in \citet{Kendall2005}).

We apply the sensitivity kernel method of Al-Attar et al. (2024) to the sea-level code of Al-Attar (2023), since it is not compatible with the evolving shoreline capability of the Han model.  We verified the code in two steps: 1. Using the same model parameters and topography and ice-sheet thickness changes, we confirmed that the difference in RSL between the models of \citet{Han2022} and \citet{Al-Attar_2023} are small ($<1\times 10^{-3}$ m) for forward model solutions; 2. We confirmed that the difference in RSL between the forward model solution and sensitivity kernel method with the \citet{Al-Attar_2023} code are very small ($<1\times 10^{-4}$ m), as reported by \citet{Al-Attar2024}.  Although we are not further developing the sensitivity kernel method in this work, this is the first systematic application of it that we are aware of to future projections of sea-level change caused by the Antarctic Ice Sheet.

To generate sensitivity kernels, we followed the procedure of \citet{Al-Attar2024}.  First, we prepared a reference dataset of initial topography and ice thickness.  Because each ISMIP6 modeling group was free to adopt different initial conditions, we calculated the average topography and ice thickness across all models.  Additionally, we calculated an ice mask for the kernel method that represents the union of the initial grounded ice across all ISMIP6 models; this ensured that changes in grounded-ice thickness from any of the ISMIP6 models were treated as a load change when applied to the kernel method, while producing an ocean area approximately appropriate for the initial ice extent from all models.  With these fields, we calculated sensitivity kernels for each of the 27 coastal cities of interest, by applying a point load smoothed over a characteristic length-scale of 1 degree as described by \citet{Al-Attar2024}.  To subsequently apply the sensitivity kernels, we integrated the sensitivity kernel for each city against specified ice thickness changes.  To account for the fixed shoreline assumption, we restricted ice thickness change to changes in height above flotation, a difference from how ice thickness change was applied to the \citet{Han2022} model.  Note that although the theory of the sensitivity kernel method demonstrates it will be of equal accuracy to a fixed-shoreline elastic fingerprint solution \citep{Al-Attar2024}, the series of small adjustments from our more complex application of the model of \citet{Han2022} (fixed shoreline, restricting load change to height above flotation, assumption of single initial condition and ocean area) introduced error, which in part motivates the comparison against the ML methods described above.

\section{Uncertainty quantification}
We construct prediction intervals for simulated sea-level change values as a means of quantifying emulator uncertainty at the 27 long-term tide gauge locations. We considered three main ways of quantifying uncertainty in this study: i) linear regression post-processing, ii) split-conformal inference, and iii) Monte Carlo dropout with latent variable sampling. We describe in more detail these three avenues for uncertainty quantification (UQ) below, and we emphasize that the use of UQ for the sensitivity kernel approach is a novel contribution to the literature.

\subsection{Linear regression prediction intervals}
As reviewed in \citet{lovegrove2023improving}, linear regression methods have been used to postprocess outputs from numerical forecasts by using the numerical model output as an input variable (i.e., covariate) for either a simple or multiple linear regression, the latter case when there are multiple covariates. In the weather forecasting literature, this has been referred to as model output statistics \citep{glahn1972use, glahn2009mos}. \citet{lovegrove2023improving} show with a global temperature forecasting example that fitting simple linear regressions yields prediction intervals for temperature that have close to nominal coverage probabilities; i.e., a (1-$\alpha$) prediction interval as derived from the simple linear regression covers the true value with probability near (1-$\alpha$) where $\alpha$ is the miscoverage rate. In addition, such an approach has been used for calibrating probabilities generated from ML models for categories via logistic regression, referred to as Platt scaling \citep{platt1999probabilistic}.

We implement this linear regression approach as follows: for a given location, we regress the simulated sea-level change values onto the emulator predictions for all of the simulation data in the validation set using a simple linear regression. This is also similar to the split-conformal method of \cite{lei2018distribution} because we use a dataset split from the original training data to fit the linear regression models, as to avoid overfitting by training both the emulator model and the simple linear model on the same data. However, we are able to improve upon conditional coverage by fitting different linear regressions for each location. Moreover, additional covariates could be included into the linear regression to improve conditional calibration. Additionally, a per-location bias is accounted for by the intercept term of the simple linear regression models.

We use the standard prediction interval from the fitted simple linear regression model:
\begin{align}
\widetilde{y}_{ij}^{test} &\pm t_{1-\alpha/2,n-2} \times  \nonumber \\
&\sqrt{ S_i^{2,val} \times \bigg( 1+\frac{1}{n} + \frac{ (\hat{y}_{ij}^{test}-\bar{\hat{y}}_{i}^{val})^2 }{ \sum_j (\hat{y}_{ij}^{val}-\bar{\hat{y}}_{i}^{val})^2 } \bigg) },
\end{align} 
where $\hat{y}_{ij}^{test}$ denotes the prediction from the emulator model for the $i^{th}$ location and $j^{th}$ test sample, $i=1,..,27$, $j=1,...,119$. Additionally, $\bar{\hat{y}}_{i}^{val}$ and $\widetilde{y}_{ij}^{test}$ indicate the location-wise mean of the prediction from the emulator model $\hat{y}_{ij}^{val}$ and the calibrated point estimate derived through a simple linear regression:
\begin{align}
\widetilde{y}_{ij}^{test}= \hat{\beta}_{i,0}^{val} + \hat{\beta}_{i,1}^{val} \times \hat{y}_{ij}^{test},
\end{align}
where $\hat{\beta}_{i,0}^{val}$ and $\hat{\beta}_{i,1}^{val}$ are derived by fitting a simple linear regression on the validation set. Also, a variance estimate for the $i^{th}$ location, $S^{2,val}_i$, is obtained by
\begin{align}
S^{2,val}_i= \frac{1}{n-2} \sum_j (y_{ij}^{val} -\widetilde{y}_{ij}^{val})^2,
\end{align}
where $y_{ij}^{val}$ represents the (simulated) sea-level change for the $i^{th}$ location and $j^{th}$ validation sample, while $\widetilde{y}_{ij}^{val}$ denotes the point estimate obtained from linear regression. Lastly, $t_{1-\alpha/2,n-2}$ denotes the 1-$\alpha/2$ quantile of a $t$ distribution with $n-2$ degree of freedom, where $n=119$ represents the sample size (i.e., the validation set size). We emphasize that we utilize the validation set not only to monitor potential overfitting but also to derive estimates for a simple linear regression. In other words, $\hat{y}_{ij}^{val}$, $\bar{\hat{y}}_{i}^{val}$, $S_i^{2,val}$, $\hat{\beta}_{i,0}^{val}$, and $\hat{\beta}_{i,1}^{val}$ are estimated based on the validation set, and these estimates are then employed to calibrate the emulator predictions.

So long as a linear model is an appropriate fit between observations (i.e., the sea-level change simulations) and emulator predictions, prediction intervals should cover with nominal probabilities. In Section \ref{result} that follows, we demonstrate empirical coverage rates on a hold-out test set of simulations. While we do not explore alternate residual distributions or spatial correlation in the residuals, we believe these items would be worth exploring in future works. Nonetheless, the split-conformal method of UQ next described easily allows for a departure from normality. 

\subsection{Split-conformal inference}
A prevalent modern technique for quantifying predictive uncertainties from ML models is that of split-conformal inference \citep{lei2018distribution}. The essence of this approach is to fit the ML model on a training dataset as usual, but use another dataset (e.g., validation) that was not used for training in order to build an empirical distribution of residuals. Then, when making a prediction for unseen test data, one uses a prediction interval $\left[ \hat{f}(\bm{x})-d_{\alpha}, \hat{f}(\bm{x})+d_{\alpha} \right]$, where $\hat{f}$ is the fitted ML model, $d_{\alpha}$ is a residual chosen from the empirical residual distribution that leads to a specified probability of coverage (i.e., $1-\alpha$), and $\bm{x}$ is a test example. We specifically implement Algorithm 2 of \cite{lei2018distribution} in computing prediction intervals in the emulator results that follow. We note that while this method has been developed primarily in the context of ML models, it is applicable to any function being fit including the physics-based sensitivity kernel approach.

This method of UQ has several advantages and disadvantages, which are summarized here. Advantages include not having to retrain the ML model as in resampling approaches (i.e., computational efficiency), no distributional assumption on residuals, and a mathematical guarantee of coverage probability meeting the nominal rate; by nominal rate of coverage, we mean the stated value of $1-\alpha$, e.g., 95\% for a 95\% prediction interval. However, this guarantee is an average over the inputs, meaning that some input cases can have coverage less than the nominal rate as long as the average meets the nominal rate. Relatedly, the interval lengths do not change for different $\bm{x}$ values, i.e., regions of input space. (By prediction interval length, we mean the upper bound of the interval minus the lower bound of the interval.) However, we account for variability in residuals across different regions in this application by using distinct residual distributions for each location, just as we consider different linear regression models for each location discussed previously. Therefore, the lengths of predictive intervals change for the different locations in our application. We note that while this method easily allows for a departure from normality of residuals, it does not explicitly account for biases in the predictions, as is done with the linear-regression intercept term described previously.

\subsection{Latent variable sampling and Monte Carlo dropout}
The previous two UQ approaches, split-conformal inference and linear regression post-processing, are quite generic and can be applied to all of the emulation methods (i.e., machine learning and physics-based sensitivity kernel) examined in this study. However we note that the CVAE ML model described in Section 3.2 is a generative ML model and implements Monte Carlo (MC) dropout \citep{gal2016dropout}, and hence adds another possible method for UQ. MC dropout is implemented in PyTorch and essentially randomly zeroes out connections between neurons in each layer with some pre-specified probability (we used 0.3). Sampling standard normal latent variables and passing them through the decoder of the CVAE also adds another source of variability in predictions. However, when we implemented this method of UQ, we found that the coverage of predictive intervals that are nominally \textbf{99+\% coverage} (i.e., minimum and maximum of samples) covered the true output only \textbf{16\%} or less of the time because the resultant predictive intervals were far too narrow. We note that if MC dropout was used without latent variable sampling, the variability in predictions would be even smaller. The summary statistics for empirical coverage rates and average interval lengths for nominally 99+\% intervals over the 27 locations are shown in Tables 1 and 2.

\begin{table}[]
\centering
\begin{tabular}{cccccc}
\hline
Min     & 1st Quartile & Median  & Mean    & 3rd Quartile & Max     \\ \hline
0.017 & 0.092 & 0.118 & 0.112 & 0.130 & 0.168 \\ \hline
\end{tabular}
\caption{Summary statistics for the empirical coverage rates over the 27 locations of interest using MC dropout for UQ. The prediction intervals here are derived from the minimum and maximum of predictive samples and should be close to .99 or greater if the MC dropout predictive distributions have well-calibrated uncertainties, since 500 samples were generated with MC dropout.}
\end{table}

\begin{table}[]
\centering
\begin{tabular}{cccccc}
\hline
Min     & 1st Quartile & Median  & Mean    & 3rd Quartile & Max     \\ \hline
0.077 & 0.084 & 0.087 & 0.086 & 0.089 & 0.095\\ \hline
\end{tabular}
\caption{Summary statistics for the average empirical prediction lengths, in \textbf{cm}, over the 27 locations of interest using MC dropout for UQ. While these lengths are very small, because of small coverage rates it can be inferred the MC dropout predictive distributions are very biased, implying poor UQ.}
\end{table}

Because these coverages are far less than the nominal rate while the predictive distributions from MC dropout are very narrow, this implies the predictive distributions from the MC dropout approach are very biased and of poor UQ quality. We disregarded this approach for UQ in the comparisons that follow.

\section{Emulator results}
\label{result}
\begin{figure}[t]
\centering
\caption{R-squared comparison between calibrated emulator predictions and actual simulated sea-level change values. R-squared values indicate that the kernel is most accurate, followed by the CVAE, and then the NN. All emulators show a high level of accuracy; see following linear plots in addition to these boxplots. The variation these display is over the 27 locations examined in this study.}

\label{results:accuracy}
\includegraphics[width=1.0\linewidth]{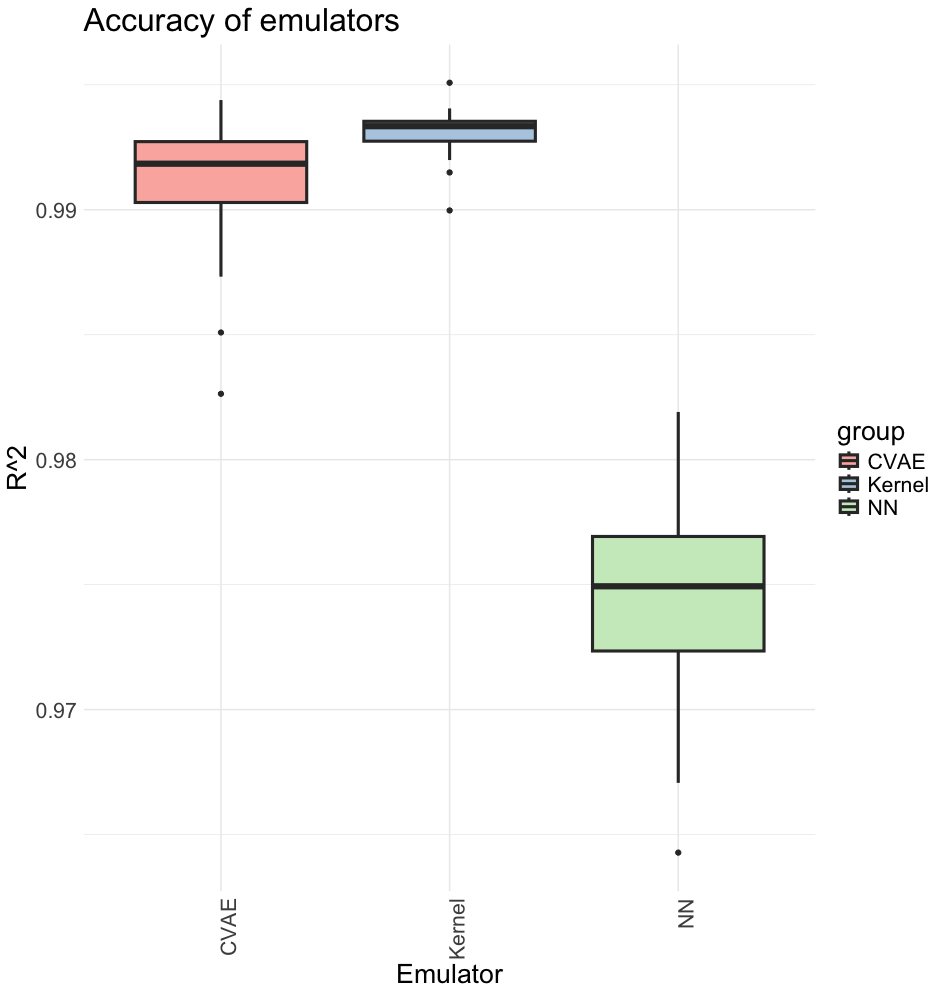}
\end{figure}
\begin{figure}[t]
\centering
\caption{Root mean squared error (RMSE) comparison between calibrated emulator predictions and actual simulated sea-level change values, in units of \textbf{cm}. RMSE values indicate similar insights as the previous figure, in that the kernel is most accurate, followed by the CVAE, and then the NN. The variation these boxplots display is over the 27 locations examined in this study. The median RMSE values are about 0.3 cm for the kernel emulator, 0.5 cm for the CVAE, and 0.7 cm for the NN emulator.}
\label{results:accuracy_RMSE}
\includegraphics[width=1.0\linewidth]{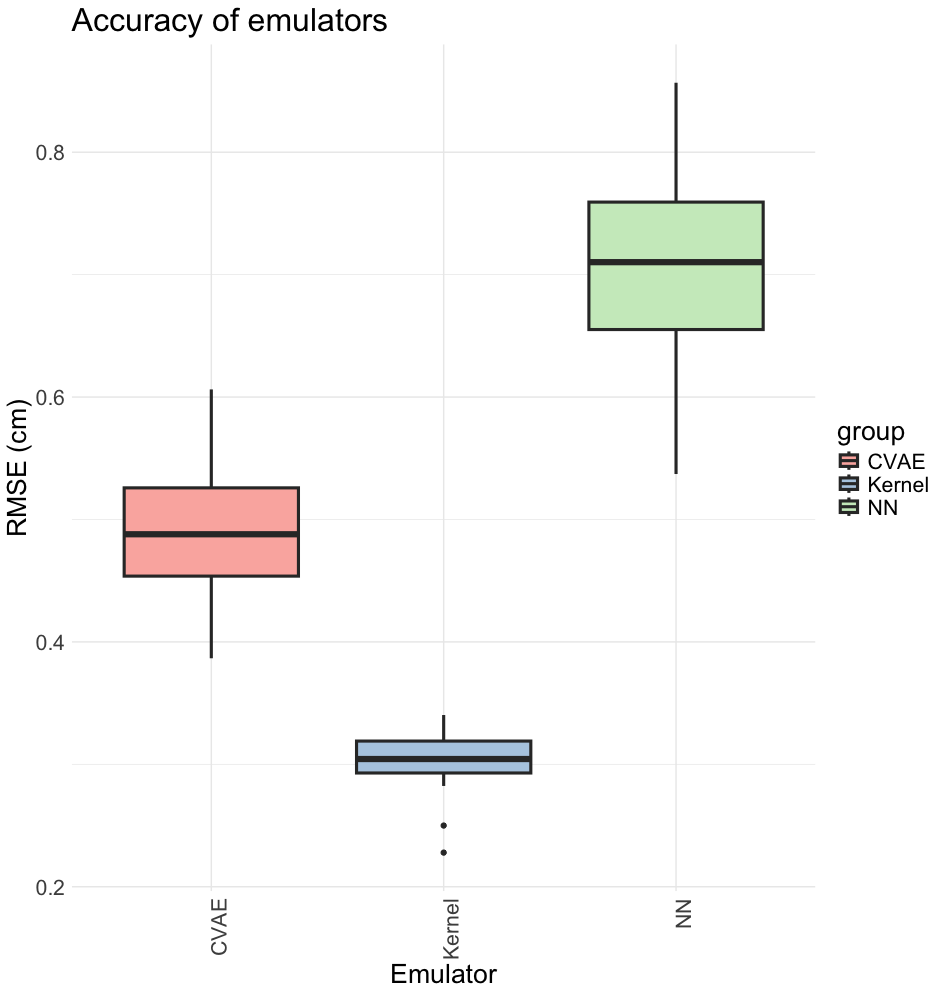}
\end{figure}
\begin{figure}[t]
\centering
\caption{Comparison of empirical prediction interval coverage rates over the three emulator methods (kernel, CVAE, and NN) and two UQ methods (linear model, subscripted as ``lm", and split-conformal inference, subscripted as ``conf"). For the 85\% and 90\% prediction intervals, the empirical coverages meet the nominal rate, excepting slight undercoverage for the kernel split-conformal intervals. For the 95\% and 99\% intervals, the split-conformal methods come closest to meeting the nominal coverage rate. Horizontal dashed line depicts the nominal coverage rate that ideally all methods are close to. The variation in these boxplots is over the 27 locations examined in this study.}
\label{results:coverage}
\includegraphics[width=1.0\linewidth]{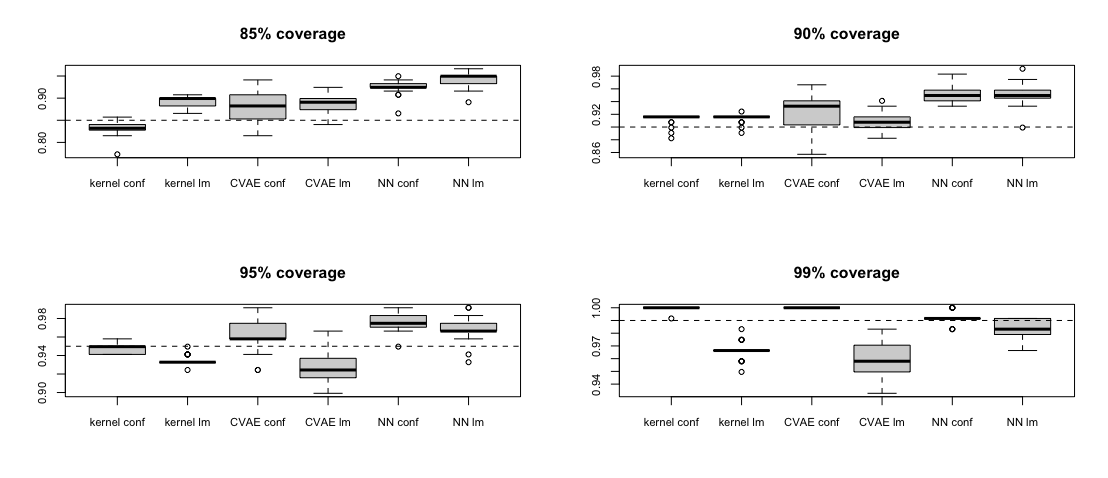}
\end{figure}
Our main results consist of a comparison of accuracy, empirical prediction interval coverage rate, and prediction interval lengths (defined as the upper bound minus the lower bound of the prediction interval) computed on the test hold-out set. We calculate Pearson's correlation coefficient of determination, $R^2$, in addition to root mean squared error (RMSE) between emulator predictions and the actual simulated sea-level change values as a measure of accuracy, though the illustrations in Figure \ref{lin_example} provide more information on accuracy. The boxplots of Figures \ref{results:accuracy} and \ref{results:accuracy_RMSE} depict the range of values for each of the emulators compared (i.e., NN, CVAE, and the sensitivity kernel method) across the 27 distinct global locations of this study. From Figures \ref{results:accuracy} and \ref{results:accuracy_RMSE}, it is evident the sensitivity kernel is most accurate, followed by the CVAE, and finally the NN. However, the CVAE is close to the sensitivity kernel, and for 6 locations achieves a larger $R^2$ value. We note that we also considered a variation where the inputs are not flattened vectors but instead based on sectors in Antarctica (see \ref{appendix_a1}), though the accuracy in that case was markedly worse than the results here, so we did not pursue it further.

We also performed paired Wilcoxon signed-rank tests to statistically compare  accuracy, measured via $R^2$ and RMSE, between the kernel and CVAE and kernel and NN emulators. The Wilcoxon signed-rank test is nonparametric and does not assume normality. We paired both $R^2$ and RMSE for each of the 27 locations when conducting these statistical tests. The p-values of all of the tests are very small, with the largest being $1.76 \times 10^{-5}$, supporting the alternative that the kernel emulator is more accurate (in terms of smaller RMSE and larger $R^2$) than either the NN or CVAE emulators. 

\begin{figure}[t]
\centering
\caption{Comparison of average interval lengths, in \textbf{meters}, comparing the three emulator methods (kernel, CVAE, and NN) and two UQ methods (linear model, subscripted as ``lm", and split-conformal inference, subscripted as ``conf"). Note that the coverage of the split-conformal method in the 99\% case comes at a substantial length tradeoff for the CVAE. The variation in these boxplots is over the 27 locations examined in this study.}
\label{results:length}
\includegraphics[width=1\linewidth]{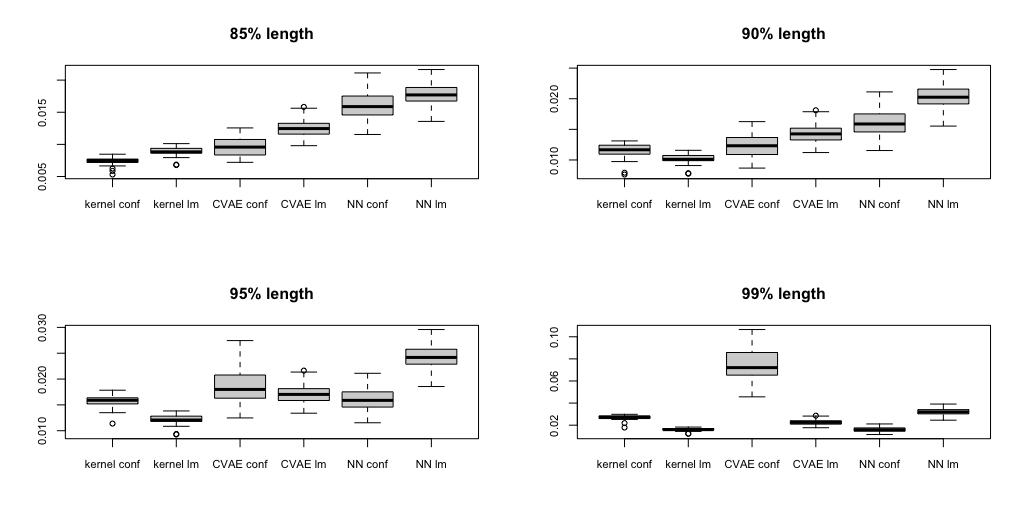}
\end{figure}

\begin{figure}[t]
\centering
\caption{Comparison of average interval scores, comparing the three emulator methods (kernel, CVAE, and NN) and two UQ methods (linear model, subscripted as ``lm", and split-conformal inference, subscripted as ``conf"). While the scores are generally better (lower score is better) for the conformal method than the linear model, this is not apparent for the CVAE and kernel methods for the 99 percent prediction intervals. The longer prediction intervals for the 99 percent conformal case can explain this behavior especially for the CVAE. The variation in these boxplots is over the 27 locations examined in this study.}
\label{results:int_score}
\includegraphics[width=1\linewidth]{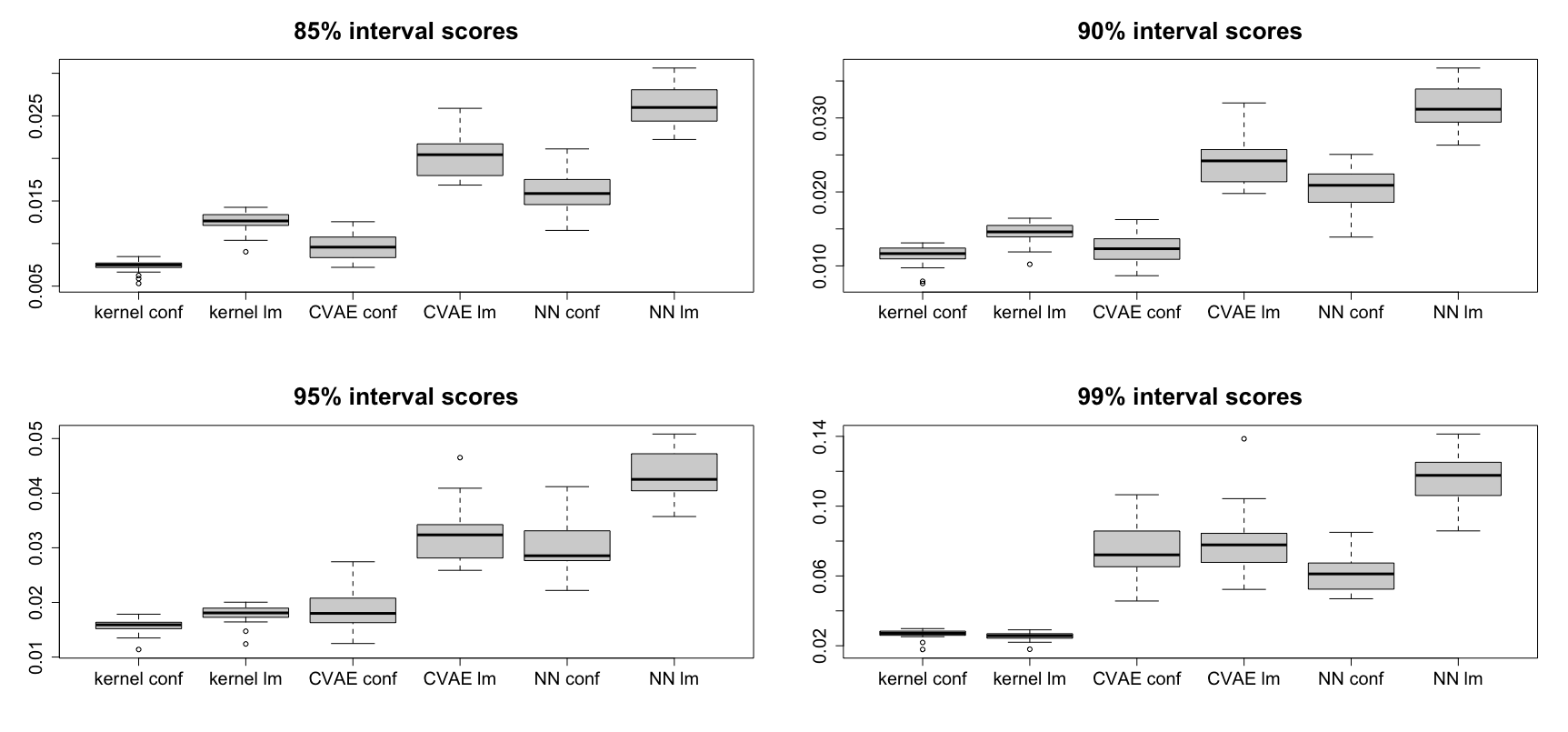}
\end{figure}

We compare the empirical coverage rates and prediction interval lengths as a means of assessing the quality of UQ for all three emulator methods and for both the split-conformal and linear-regression methods of UQ. We show those for nominal miscoverage values of $\alpha = 0.01, 0.05, 0.10$, and $0.15$ in Figures \ref{results:coverage} and \ref{results:length}. Our findings suggest that the best coverages are achieved by the split-conformal method, though the lengths of the intervals, particularly for $\alpha = 0.01$, may be too large to be of practical value in applications. The lengths of predictive intervals for the linear regression method are far more useful (in terms of length) for smaller values of $\alpha$ but these intervals undercover especially for $\alpha = 0.01$. In addition to empirical coverage rates and prediction interval lengths, we also compute the average proper interval score \citep{gneiting2007strictly} per location and display these in Figure \ref{results:int_score}; the findings here are similar to the analysis of coverage and length, in that the split-conformal scores are general lower (i.e., better), but in the 99\% case split-conformal and the linear model approach are nearly tied for the kernel and CVAE methods, and slightly larger for the NN method using the linear model approach. As is reviewed more in \cite{gneiting2007strictly}, the proper interval scoring rule combines empirical coverage and interval length into a single metric, which penalizes prediction interval length as well as empirical miscoverage relative to the nominal rate. For example, if the true value being predicted is a fixed distance away from the lower bound of a prediction interval, this will be penalized more for a higher nominal coverage rate.

Figure \ref{lin_example} shows an illustration depicting the linear-regression results used for UQ for three locations that represent a wide variation of RSL change amongst the 27 cities: Midway, Montevideo, and Dunedin. The three sites chosen were selected as they represent among the widest variation in RSL of the cities sampled (Figure \ref{fig:sim_example}). Simulated sea-level change (i.e., the prediction target) is on the $y$-axis and emulator predictions are on the $x$-axis. Each column represents the three emulator methods in this study: NN, CVAE, and the kernel method. Additionally, we plot histograms of residuals derived for use in split-conformal inference in Figure \ref{results:conformal_example}. These residuals suggest that the scale of errors is smallest for the kernel method, further corroborating the claim that it is the most accurate of the emulation methods compared.

\begin{figure}[t]
\centering
\caption{Scatterplots comparing sea-level change simulation targets (y-axis) with emulator predictions (x-axis) on the test set for Dunedin, Montevideo, and Midway, in \textbf{meters}. Solid line indicates when predictions equal simulations, i.e. a perfect fit. The kernel method is most accurate, followed by the CVAE, and then NN. The kernel method also extrapolates best -- it is most accurate for the most extreme point for all three locations.}
\label{lin_example}
\includegraphics[width=.85\linewidth]{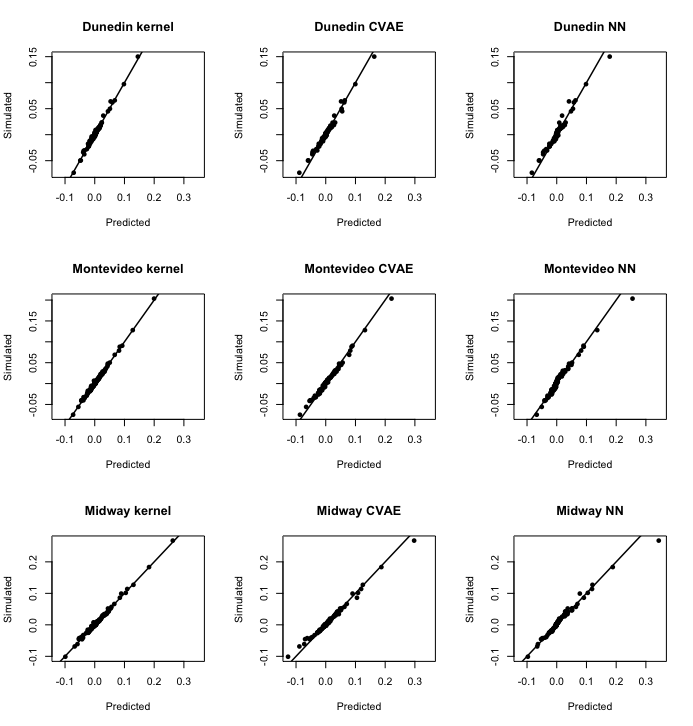}
\end{figure}
\begin{figure}[t]
\centering
\caption{Empirical distributions of residuals in \textbf{meters} used for split-conformal inference illustrated with histograms at the same three locations: Dunedin, Montevideo, and Midway, which showcase a wide spectrum of RSL variation. Consistent with the previously shown scatterplots, the residuals are smallest in magnitude for the kernel method. The residual distributions are quite similar for the CVAE and NN, despite that the CVAE appears to have substantially higher R-squared values.}
\label{results:conformal_example}
\includegraphics[width=.85\linewidth]{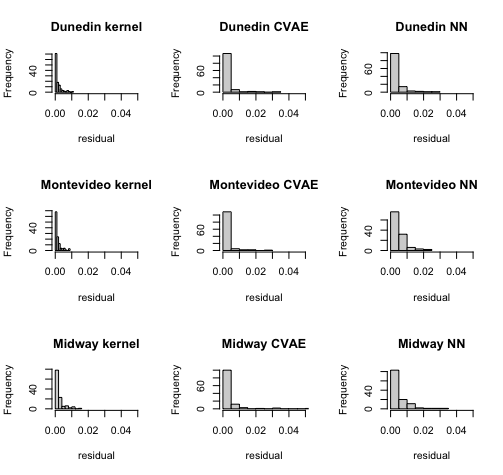}
\end{figure}

We additionally assess the runtime for generating predictions with each of the ML models by performing a single prediction task 1,000 times. All 1,000 NN predictions completed in less than 0.25 seconds and all but one of the CVAE runs completed in less than 0.375 seconds. We consider that a single prediction is complete when the model provides the predicted values for all 27 locations. For the CVAE model, the runtime is measured based on a single sample drawn from the assumed latent distribution.  We also observe that the additional calibration step does not contribute to a significant increase in runtime ($\approx$ 0.0006 seconds). Training each of the ML emulators takes on the order of a few hours on top of the overhead of ~21 hours (17 minutes per a single run, times 71 runs) for running the sea-level fingerprint calculations in serial. In contrast, the kernel method's runtime is due to a matrix-vector multiplication and requires only a single forward solve for generating the sensitivity kernel and thus substantially much less overhead than the NN and CVAE for training. The kernel method takes a mean and median of 5.2 seconds to generate emulated sea-level change values for all 27 locations of interest for all 17 time levels; thus taking 0.3 seconds to make a prediction at all 27 locations, which is close to the CVAE and NN runtime. Additionally, the kernel method took 33 seconds to generate the kernels for the 27 cities.
Given that the full solve takes about 60 seconds per time level, all emulation methods afford a speed up of more than 100 times over running the full simulation for generating a sea-level fingerprint; we stress, however, that the kernel-generation overhead for the kernel method is substantially less than the training overhead for the ML methods.

In Table 3, we summarize the computational costs for the emulators for three steps: 1) generating the sea-level fingerprints via the numerical sea-level solver, training the emulators on the training data, and evaluating the emulators to make a prediction for the 27 locations of interest. 
\begin{table}[]
\centering
\begin{tabular}{cccccc}
\hline
Emulator     & Data generation (min) & Emulator fitting (min)   & Time for predictions (s)    \\ \hline
Sensitivity kernels & 1  & NA & 0.31 \\ \hline
NN & 969 & 121 & 0.25 \\ \hline
CVAE & 969 & 286 & 0.375  \\ \hline
\end{tabular}
\caption{Summary of computational runtime comparisons for generating sea-level fingerprints used for building emulators, training the emulators, and making predictions for 27 locations of interest, after the emulators have been trained. Note that for the kernel method, the kernel for each location is created upon running the sea-level solver such that no additional training is needed (in contrast to the ML methods). The ML models were trained using a shared graphics processing unit (GPU) partition available from the Darwin \citep{osti_1441285} cluster at Los Alamos National Laboratory.}
\end{table}

\section{Application of emulators for a Monte Carlo ensemble}
We demonstrate a potential use of the sea-level change emulators for generating distributions of simulated sea-level change outputs for locations across the globe, in order to show how such an emulator could be used to produce RSL distributions from future Antarctic ice-sheet simulations. The most straightforward Monte Carlo approach to do so is to sample many parameters from their respective distributions, run an ice-sheet simulator under the sampled parameters, and subsequently run the resultant ice-sheet states through the sea-level change emulator in order to generate a distribution of sea-level change values. For our purposes, running an ice-sheet simulator iteratively is prohibitively expensive and so we instead construct a Monte Carlo ensemble of ice-sheet states using the ISMIP6 ensemble of Antarctica simulations.

\begin{figure}[t]
\centering
\caption{Probability density plots of emulated RSL change for Dunedin, Montevideo, and Midway during year 2100 with high-emissions scenario, produced with the kernel emulator, which is most accurate of all emulators compared previously. We expect Midway to have the heaviest right tail given its location in the Pacific and Dunedin to have the lightest right tail. Both aspects are evidenced here.  Note that these are densities for simulated RSL change and are an example of the application of the emulator. RSL units in \textbf{meters}.}
\label{kernel_emulated_dist}
\includegraphics[width=.75\linewidth]{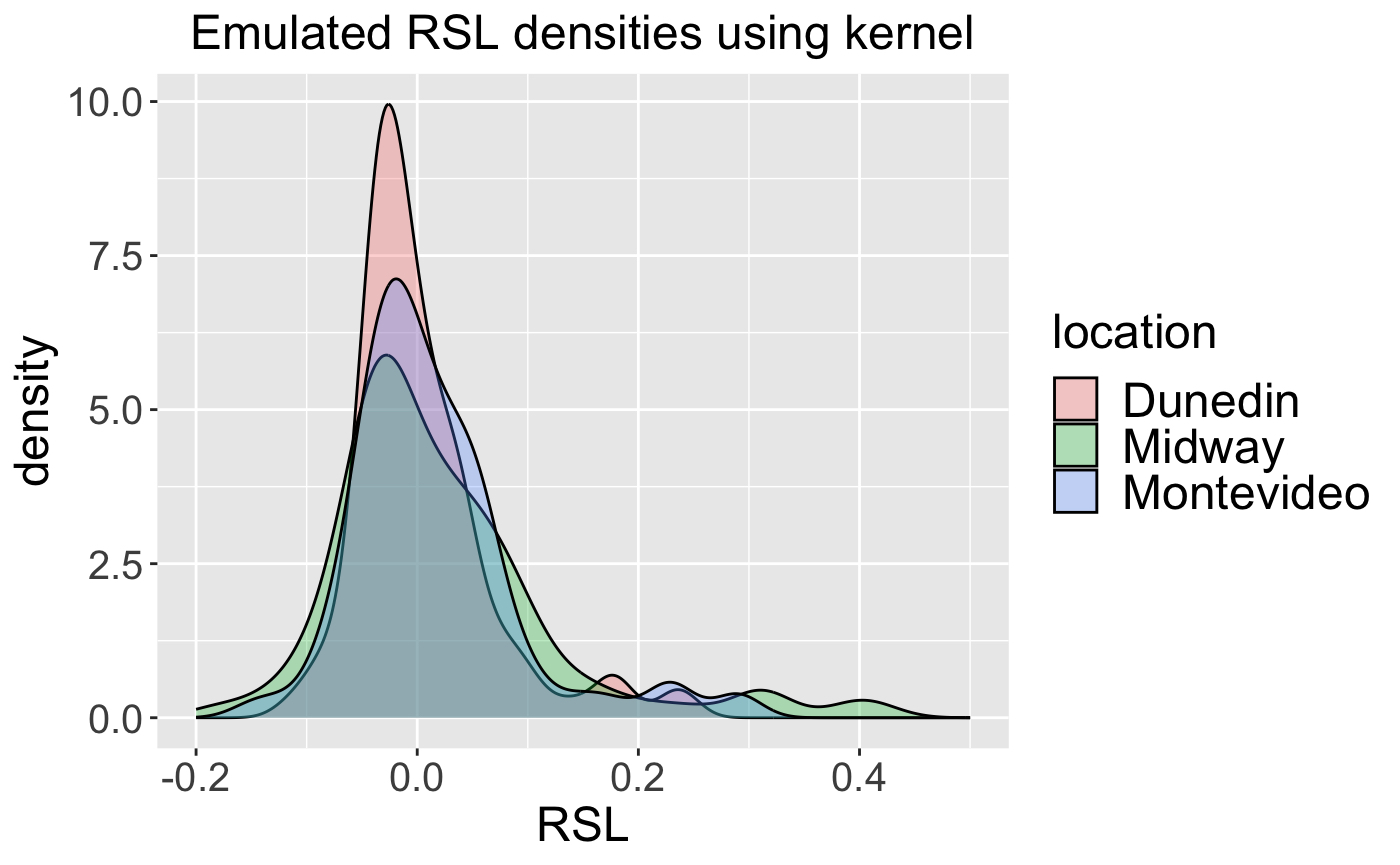}
\end{figure}

\begin{figure}[t]
\centering
\caption{Probability density plots of emulated RSL change for Dunedin, Montevideo, and Midway during year 2100 with high-emissions scenario, produced with the NN emulator, which is least accurate of all emulators compared previously. The right tails of these densities are similar to the previous plots in that Midway has the heaviest right tail whereas Dunedin has the lightest right tail, though the difference is most pronounced in the previous densities -- which are more reliable due to the kernel method's better extrapolation ability. Note that these are densities for simulated RSL change and are an example of the application of the emulator. RSL units are in \textbf{meters}.}
\label{NN_emulated_dist}
\includegraphics[width=.75\linewidth]{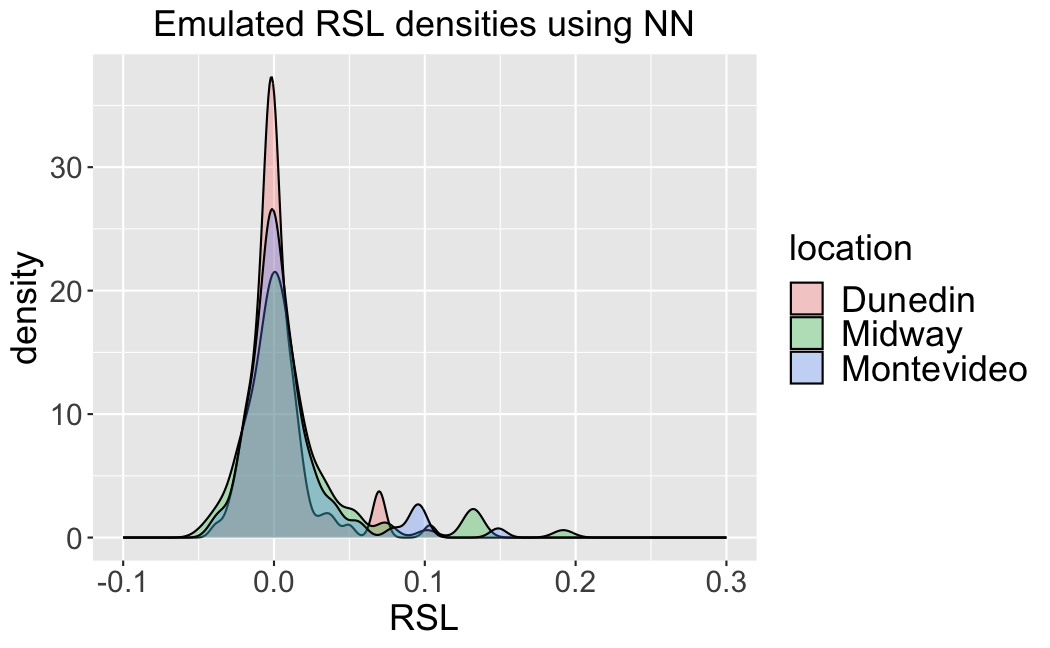}
\end{figure}

Our goal in the demonstration is to use the kernel and NN emulators to produce a distribution of sea-level change values for the 27 long-term tide gauge sites. For this example, we focus on the year 2100 and high-emissions scenarios, and cull 48 instances from the ISMIP6 training data matching these criteria. We performed a principal-components analysis on these 48 instances, retaining the top 32 which explained 99\% of the variation in the year 2100, high-emissions simulations. We then bootstrapped the principal-component weights 100 times in order to randomly generate a Monte Carlo ensemble of Antarctic ice-sheet states representative of the year 2100 and high emissions. Alternative dimension reduction techniques could be explored, though a principal-components approach was used as a practical method to generate a Monte Carlo ensemble. These 100 randomly generated ice-sheet states were then fed into the kernel and neural-network emulators. We illustrate emulated densities of sea-level change using the Monte Carlo simulations in Figures \ref{kernel_emulated_dist} and \ref{NN_emulated_dist}. We emphasize that these distributions are just meant as a demonstration of the application of the emulators.

As in Section \ref{result}, the three locations were selected because they are indicative of the widest variation in RSL of the cities sampled (Figure \ref{fig:sim_example}).  The emulated RSL derived from the synthetic ice-sheet states span a similar range of sea-level change as the GMSL estimates from the ISMIP6 ensemble from which they were derived; \citet{Seroussi2020} report mean sea-level contribution of about 0.025 m with a standard deviation of 0.049 m for the high emissions scenarios.  \citet{Seroussi2020} also demonstrate a longer tail of positive GMSL change in the ensemble, which our distributions also show (Figures \ref{kernel_emulated_dist} and \ref{NN_emulated_dist}).  We consider this consistent given we did not use all ISMIP6 samples, our synthetic ice-sheet states are not guaranteed to have the same statistical properties as the ISMIP6 ensemble, we are considering RSL rather than GMSL, and these results come from the kernel and NN emulators.

While there are clear differences in the densities of RSL projections for the three locations, they share many similarities, despite being from different parts of the Earth.  
The mode of each density close to zero is a result of our synthetic ice-sheet states generating many samples with small net mass loss from AIS.  However, the tails of the distributions exhibit significant differences.
Positive RSL values have lower magnitude for the two Southern Hemisphere locations (Dunedin and Montevideo), consistent with the  fingerprint of decreased RSL rise closer to AIS (Figure. \ref{fig:sim_example}).  This has the effect of compressing the range of the density functions.  The normalized RSL changes at southern hemisphere locations is more sensitive to the spatial pattern of AIS mass change \citep{Cederberg2023, Roffman2023}, but due to the smaller magnitude of RSL change there, also have smaller standard deviations \citep{Cederberg2023}.  While more detailed evaluation of these synthetic distributions is outside the scope of this work, this example shows the potential utility of emulators for producing RSL estimates rapidly from very large numbers of ice-sheet simulations.

\section{Discussion}

\subsection{Comparison of kernel and machine-learning emulators}

The derivation of the sensitivity kernel method is based on a physics approach that makes use of reciprocity theorems as is described in Section 4. As such, because this physics-based emulator is tailored to the problem of emulating sea-level change and makes use of symmetry specific to this system on timescales where elastic effects are dominant, it is perhaps unsurprising that it performs better than the generic ML approaches compared. In contrast, the ML approaches do not have any additional information about the underlying physics besides what is represented in the training data set input-output pairs. The sensitivity kernel emulator appears to perform better than the ML emulators based on statistical tests, though the improvement in accuracy is not, for instance, several orders of magnitude. Nonetheless, the kernel method does extrapolate better for extreme input values than the ML methods. All three emulators evaluate predictions in a comparable time frame (i.e., less than 0.5 seconds), but the overhead for generating training simulations for training the ML methods is more than 20 hours whereas the kernel training just take a forward solve (i.e., a minute). 

\subsection{Emulator limitations and future work}

The most similar ML application to sea-level change is the ``Graph Neural Network Based EmulatOR for Glacial Isostatic Adjustment'' by \citet{lin2023georgia}.  This work used a similar sea-level model (equivalently called a glacial isostatic adjustment model) to generate a training dataset of RSL change from a range of ice-sheet mass change trajectories, which was used to construct a convolutional neural network (CNN).  A key difference is that their training dataset used paleo ice-sheet changes over tens of thousands of years, targeting a different application.  They also utilized a graph-based spherical CNN, which allows convolution operations on a spherical manifold appropriate for representing the entire Earth.  That is a promising technique for whole-Earth RSL emulation, and their emulator presumably could be trained on future AIS projections (and other sea-level contributors).  For quantifying uncertainty, \citet{lin2023georgia} utilized an ensemble learning approach, training 30 versions of their emulator. However, their 99\% prediction intervals cover the true values reportedly 94.5\% of the time, suggesting that actual coverage rates are less than nominal coverage rates. Additionally, such an ensemble approach has the increased computational burden of having to refit the neural network multiple times during training. 


While this study focuses on 27 specific locations, it is of interest to build a RSL emulator that produces spatial output for the entire globe, where the dimension of the output $\bm{y}$ significantly increases thereby introducing more complexity. Our preliminary work on this global spatial emulation problem has yielded varying quality of emulated outputs. Better performance could be achieved by exploring modern NN architectures, including CNNs such as U-Net \citep{ronneberger2015u} and the transformer \citep{vaswani2017attention}, which could prove beneficial by effectively capturing spatial dependencies but come at the potential cost of requiring a prohibitively large training dataset of simulations. An additional challenge of emulating the entire globe is ensuring spherical consistency of the emulated RSL field.  While the native output of the sea-level model is a structured latitude-longitude grid, treating that as two-dimensional space would ignore the fact that the longitude points are periodic and lines of latitude  converge at the poles.  Using a spherical representation of the output space \citep{lin2023georgia} or emulating the spherical harmonic coefficients directly are avenues for addressing this issue.

Beyond the small dimension of our output space, there are ways in which the training data could be improved.  The sample of the ISMIP6 ensemble we used here was relatively small and exhibited a fairly limited range of potential AIS states.  Recent single-model ice-sheet model ensembles have produced hundreds \citep[e.g.][]{DeConto2021, Berdahl2023, Coulon2024, Aschwanden2022} or thousands \citep[e.g.][]{Bulthuis2019} of model realizations, and by exploring parameter uncertainty produce a large spread of potential ice-sheet states.  Utilizing more expansive training data would improve both the accuracy of the emulator as well as its applicability, but with the trade-off of increased cost of generating the RSL training data with the sea-level model.  A more selective approach to choosing training data could also make the process more efficient --- because many of the ensemble members had mass change close to zero, the emulators may have been focusing on behavior of little interest.

Other areas for improvement include extending the RSL emulation to additional source of sea-level change (Greenland Ice Sheet, mountain glacier and icecaps, terrestrial water storage) and considering more complex Earth structure.  As mentioned above, lateral variations in mantle viscosity and lithospheric thickness \citep{Barletta2018, Nield2014} have significant impacts on RSL derived from AIS \citep{Pan2021, Powell2022}.  Models accounting for 3-D viscoelastic Earth structure are significantly more computationally expensive to run \citep[e.g., factor of $\mathcal{O}(10^4)$,][]{Love2023}, which would greatly increase the value of developing fast emulators.  \citet{Love2023} demonstrated potential for ML to capture the effects of lateral variations in Earth structure at a fraction of the computational cost.

\subsection{Recommendations}
Our results point to the sensitivity-kernel emulator being the most accurate and requiring the least computational overhead to generate. The neural network methods we had compared against, as well as the others discussed, have shown that NNs are also viable options for sea-level change emulators. 
As described above, the short timescale considered here allowed us to use an elastic Earth structure for the RSL predictions.  Over longer timescales where viscoelastic Earth deformation is relevant, the elastic kernel method will no longer be appropriate and the ML techniques may be more accurate.
We also note than an alternative approach to the sensitivity-kernel method would be exploring forward model elastic fingerprints with reduced dimensionality of the input space.  This could either be reduced spherical harmonic degree and order or representing the ice-sheet mass change as uniform within prescribed sectors.  Sector-based fingerprints could be generated and subsequently scaled by mass change in each sector and then added together to generate RSL maps, exploiting the linear nature of elastic fingerprints.  We did not explore this option because the sensitivity-kernel approach allowed us to retain the complete structure of the AIS thickness change field, but for applications where the global pattern of RSL is of interest, as opposed to a small number of point locations, a sector-based fingerprint approach could be advantageous.

Despite their substantial predictive capabilities, designing the architectures of both the NN and CVAE, including parameters like the number of layers ($L$), the quantity of units ($d_{\ell}$), the selection of activation functions, and so on, may not be straightforward and can be application specific. The process of fitting multiple models with varying architectures to pinpoint an optimal design can be somewhat \textit{ad-hoc} and time consuming. Alternatively, NAS has proven to be effective in automating the architecture search \citep{nas}. Therefore, identifying appropriate model structure for sea-level change emulators through NAS could be helpful in constructing ML-based emulators.
Moreover, software such as PyTorch \citep{pytorch}, among others, has enabled the straightforward use of a variety of neural network architectures and different types of loss functions, creating many possibilities for the scientist.

We have demonstrated the utility of two UQ methods for augmenting emulator predictions with reliable uncertainty estimates: simple linear regression and split-conformal inference. The simple linear regression method of performing UQ on top of emulator output shares elements from the weather forecasting literature \citep{glahn1972use, glahn2009mos, lovegrove2023improving}, as well as split-conformal inference \citep{lei2018distribution}. Some advantages of the approach are that that it is relatively inexpensive to implement after the initial emulator has been fit and can be used irrespective of the specific emulator and/or loss function used; it is therefore recommended as an additional approach to UQ for ML predictions in addition to split-conformal inference. In contrast, Monte Carlo dropout appears to produce overly narrow intervals for the neural network predictions in this study.

\section{Conclusion}
We have demonstrated the efficacy of sensitivity-kernel based methods in emulating simulated sea-level change  at a substantially reduced computational cost and high accuracy. In comparison, the NN and CVAE models were close though not as accurate in emulating sea-level change. Employing UQ via split-conformal inference and a linear-regression postprocessing technique produced prediction intervals that are generally well calibrated.  

We have  highlighted the use of both split-conformal inference and a simple linear regression method for deriving statistically valid uncertainty quantification using (nonlinear) ML predictions. The linear regression method of calibrating uncertainties is rooted in the weather forecasting literature \citep{glahn1972use, glahn2009mos, lovegrove2023improving} and also shares commonalities with split-conformal inference \citep{lei2018distribution}. The method is analogous to Platt scaling \citep{platt1999probabilistic}, used for calibrating probabilities of discrete, as opposed to continuous, events. Both UQ approaches (linear regression postprocessing and split-conformal inference) in this work are straightforward to implement, computationally inexpensive, and have yielded well-calibrated prediction intervals of simulated sea-level change that are generally of a useful length, meaning that their scale is not too large -- for instance 1.6 cm for 95 \%split-conformal prediction intervals using the kernel method. Moreover, UQ is a novel contribution to the application of sensitivity kernels for SLR emulation. 

Finally, we have demonstrated an application of the kernel-based and NN emulators to a large sample of synthetic ice-sheet states and have shown their promise for producing probabilistic projections of RSL at different locations. Further work is necessary to develop an RSL emulator applicable to the whole Earth and that accounts for important lateral variations in Earth structure.  However, the emulators constructed and compared in this work show promise as a tool to accelerate workflows for producing sea-level projections that are fast, reliable, and with quantifiable uncertainty. We expect the ML emulators to especially be valuable when longer time-scale (e.g., 2300) RSL projections are needed, where viscoelastic rheology renders the use of a kernel-based method less straightforward to implement, and the inherent nonlinearites present in that regime motivate greater use of ML models such as the neural networks presented in this work. 

\appendix
\begin{figure}[t]
\centering
\caption{A map of sectors in Antarctica from the ISMIP6 basins where numbers represent distinct basins. This figure shows the sectors we used as an alternative approach to the flattened input vector of the neural networks. While the results of the sector-based inputs were not nearly as accurate as the flattened-vector inputs, these sectors are displayed to illustrate this alternative approach to the reader.} 
\label{sectors}
\includegraphics[width=0.8\linewidth]{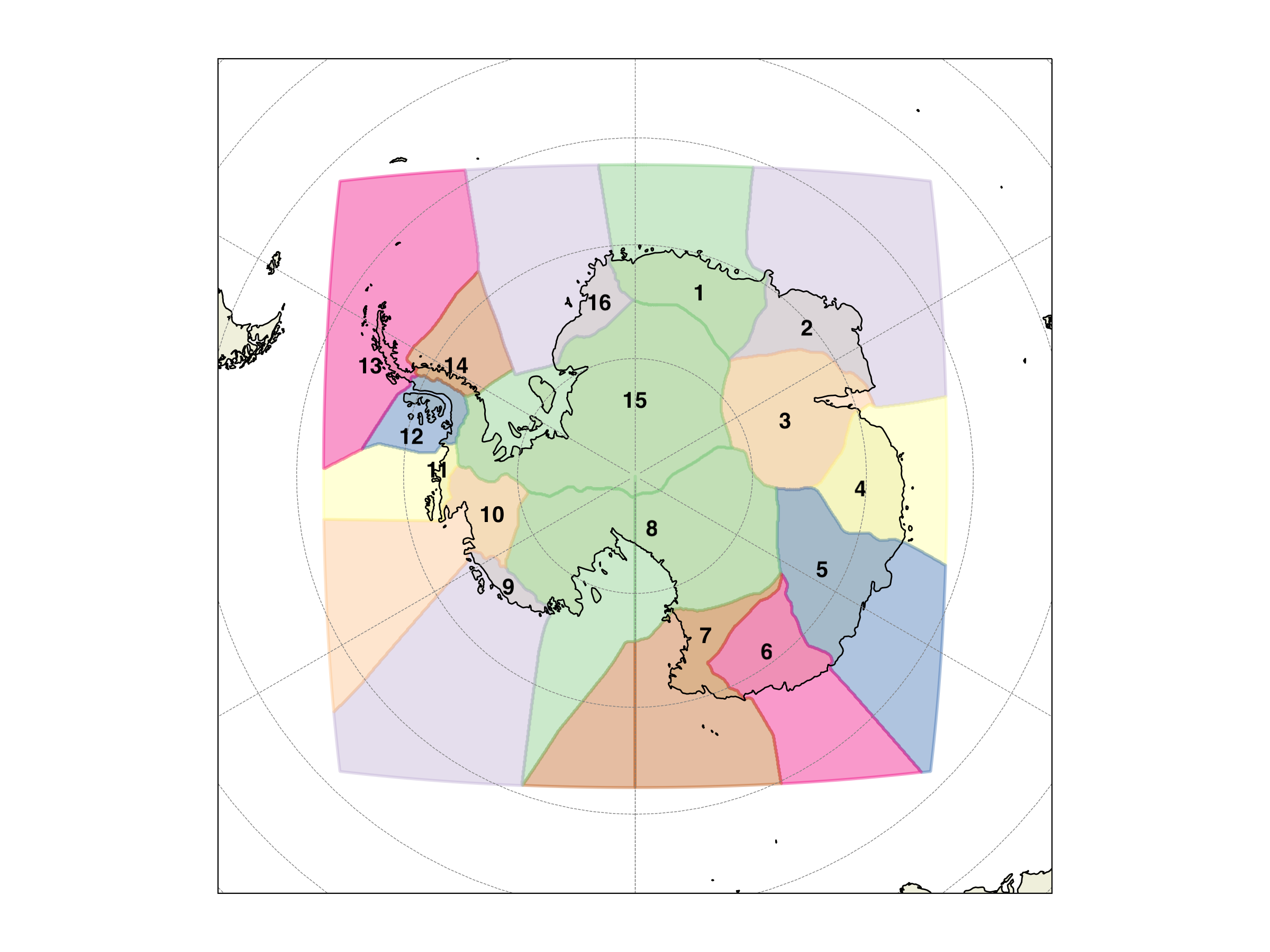}
\end{figure}

\section{Using sectors in Antarctica}
\label{appendix_a1}
We examine the prediction performance of ML models using the mean ice-thickness of geographic sectors across Antarctica as inputs, similar to the work of \cite{VanKatwyk2023}. The model structure is the same as described in Section \ref{sec3} except with using the mean thickness value for each of the 18 regions as inputs (Figure \ref{sectors}). We found that the prediction accuracy, however, was substantially than using the entire flattened vector of ice thickness changes as input (i.e., $R^2$ less than 0.1 in all cases), and so did not pursue the sector-based approach further.

\section{Alternative emulators considered}
Gaussian processes have been a common choice to build emulators for physical applications \citep{higdon2008computer, gu2018robust} and hence serve as a natural comparison to NN and CVAE emulators. They are primarily used because of closed-form multivariate normal distributions conditioning on training data; see \cite{gramacy2020surrogates} for exact formulations.

For the baseline GP emulator, we use a GP that maps from a reduced dimensional input space to the sea-level change at each of the 27 long-term tide gauge locations. The dimension-reduction technique we adopt in this work is discussed subsequently. The GP uses a squared-exponential kernel \citep{williams2006gaussian, gramacy2020surrogates} whose parameters, including a nugget term (i.e., noise term), are fit with a statistical technique known as profile likelihood maximization, the mathematical steps of which are described in Chapter 5 of \cite{gramacy2020surrogates}. Our implementation of profile likelihood was coded in the R programming language. While there exist some recent variants of GPs such as deep Gaussian processes (e.g., \citet{sauer2023vecchia}) that are capable of handling nonstationarity, we use a stationary GP with a squared-exponential covariance function only to serve as a baseline comparison.

For baseline comparisons, we use the ``RandomForestRegressor" function in a package ``scikit-learn" \citep{scikitlearn} to train an RF in Python 3.9.17. We use 100 decision trees (default) and mean squared error (MSE) as the loss function for this baseline RF. As is the case for the GP baseline, an RF is trained for each of the 27 long-term tide gauge locations using the reduced dimensional input. 

While the NN and CVAE models take as input an entire vector of ice thickness changes across Antarctica and perform dimension reduction within their respective architectures, the GP and RF models use a reduced-dimension space as input. Dimension reduction was performed by projecting input vectors onto a reduced-rank principal-component basis; 250 principal components were used, which represent almost 98.7\% of the total variation of the training set inputs. This technique is standard for emulating high-dimensional computer simulator output (e.g., \cite{higdon2008computer}, \cite{hooten2011assessing}, \cite{gopalan2022higher}), but in this work we use dimension reduction on the space of inputs for the GP and RF emulators, which is likely required as the dimension of inputs grows substantially --  the input vector length is on the order of 100,000 whereas there are order 1000 data vectors. For instance, computing the distance matrix needed for a GP model is an $O(n^2m)$ operation when there are $n$ length $m$ vectors, which for $m >> n$ can be more substantial than the $O(n^3)$ inverse and determinant calculations needed for GPs.

\section*{Open Research}
The data and code for emulators based on ML models are preserved in \citet{yoo_2025_} and developed at \url{https://github.com/msyoostat/RSL}.

The ISMIP6 ensemble dataset was retrieved from the ISMIP6 archive (\url{https://theghub.org/dataset-listing}). Scripts for preparing the ISMIP6 ensemble data for use in this study are preserved in \citet{matt_hoffman_2025_14868920} and are developed at \url{https://github.com/SLE-E3SM/ISMIP6_processing}.
The 1DSeaLevelModel\_FWTW sea-level model used for generating the elastic sea-level fingerprints is preserved in \citet{holly_han_2025_14868955} and developed at \url{https://github.com/MALI-Dev/1DSeaLevelModel_FWTW}.
The version of SLReciprocityGJI used for generating sensitivity kernels is slightly modified from \citep{Al-Attar_2023}.  The version used in this study uses modified constants to be consistent with 1DSeaLevelModel\_FWTW and includes additional pre and post processing scripts for the analysis workflow in this study.  The modified version used in this study is preserved in \citet{david_al_attar_2025_14868996} and developed at \url{https://github.com/matthewhoffman/SLReciprocityGJI}.

\section*{Acknowledgments}
The authors are grateful for support to MY from the 2023 Los Alamos National Laboratory Applied Machine Learning Research Fellowship.
MJH, GG, SC, HKH, and TH were supported by the DOE Office of Science Early Career Research program.
SC was also supported by the Laboratory Directed Research and Development program of Los Alamos National Laboratory under project number 20210952PRD3.
GG was additionally supported by the Verification and Validation subprogram of the Advanced Simulation and Computing Program at the Los Alamos National Laboratory.
MY and CKW are grateful for support from the University of Maine (NIH) subcontract 00082247 for work performed at the University of Missouri.
This research used resources provided by the Los Alamos National Laboratory Institutional Computing Program, which is supported by the U.S. Department of Energy National Nuclear Security Administration under Contract No. 89233218CNA000001.
Simulations were performed on machines at the National Energy Research Scientific Computing Center, a U.S. Department of Energy Office of Science User Facility located at Lawrence Berkeley National Laboratory, operated under Contract No. DE-AC02-05CH11231 using NERSC award using NERSC award ERCAP0028051.


%
%


\bibliographystyle{apalike}  
\bibliography{references}  

%
%
%
%
%

\end{document}